%% file: 1_maintext.tex
\documentclass[%
 reprint,
superscriptaddress,
 amsmath,amssymb,
 aps,
floatfix,
]{revtex4-2}

\usepackage{graphicx}
\usepackage{dcolumn}
\usepackage{bm}
\usepackage{upgreek}
\usepackage{float}

\begin{document}

\preprint{APS/123-QED}

\title{Temperature-tunable spin-wave refraction using superconducting control elements} 
\author{Pim H. Vree} \affiliation{Department of Quantum Nanoscience, Kavli Institute of Nanoscience, Delft University of Technology, 2628 CJ Delft, The Netherlands} \author{Merel A. Bouma} \affiliation{Department of Quantum Nanoscience, Kavli Institute of Nanoscience, Delft University of Technology, 2628 CJ Delft, The Netherlands} \author{Michael Borst} \affiliation{Department of Quantum Nanoscience, Kavli Institute of Nanoscience, Delft University of Technology, 2628 CJ Delft, The Netherlands} \author{Tomas T. Osterholt} \affiliation{Institute for Theoretical Physics, Utrecht University, 3584 CC Utrecht, The Netherlands} \author{Rembert A. Duine} \affiliation{Institute for Theoretical Physics, Utrecht University, 3584 CC Utrecht, The Netherlands} \affiliation{Department of Applied Physics, Eindhoven University of Technology, 5600 MB Eindhoven, The Netherlands} \author{Toeno van der Sar} \email{t.vandersar@tudelft.nl} \affiliation{Department of Quantum Nanoscience, Kavli Institute of Nanoscience, Delft University of Technology, 2628 CJ Delft, The Netherlands}

\
\date{\today}

\begin{abstract}
\input{sw_sc_abstract.tex}
\end{abstract}

\maketitle

\section{\label{sec:introduction}Introduction}
Spin waves are collective excitations of the coupled spins in magnetic materials. They play an important role in the temperature dependence of magnetization and are promising for realizing wave-based signal control in information devices \cite{barman_2021_2021,pirro_advances_2021,csaba_perspectives_2017,yuan_spin-wave-based_2023, candido_predicted_2020, trifunovic_long-distance_2013,chumak_advances_2022,chumak_magnon_2015, yuan_quantum_2022}. Through exchange and dipolar coupling, spin waves form strongly interacting systems with a rich, many-body physics that has sparked intense theoretical interest. Combined with non-reciprocal transport and low intrinsic damping, spin waves may enable information devices with functionalities and efficiencies beyond those based on other excitations \cite{barman_2021_2021,pirro_advances_2021,chumak_advances_2022,chumak_magnon_2015, yuan_quantum_2022,flebus_two-fluid_2016}. 

A key goal of spin-wave research is to realize devices with optics-like, wave-based functionalities that, instead of operating at the 100’s of terahertz frequencies of light, operate in the microwave/gigahertz regime important for applications such as wireless communication or (quantum) computing \cite{barman_2021_2021,pirro_advances_2021,yuan_quantum_2022}. An important advantage of using spin waves for microwave control is the $\sim$1000-fold-reduced wavelength with respect to electromagnetic waves of the same frequency in electronic circuits. This reduction enables microwave control in nano-to-microscale devices \cite{pirro_advances_2021, chumak_advances_2022,wang_reconfigurable_2018,pirro_interference_2011,wang_magnonic_2020}. 

In analogy with their visible-light counterparts, spin-wave optical elements may be realized by controlling the refractive index governing the spin-wave propagation. Pioneering experiments have shown that changing the magnetic film thickness enables modulating the spin-wave refractive index with demonstrated applications such as Fabry-Perot resonators \cite{stigloher_snells_2016,qin_nanoscale_2021}. Furthermore, periodic structures of magnetic metals on magnetic insulators have enabled spin-wave gratings, crystals and non-reciprocal transport devices \cite{qin_nanoscale_2021,yu_omnidirectional_2013,chen_excitation_2019,chen_strong_2018,qin_low-loss_2018,chumak_magnonic_2017}. 

An emerging method that could enable low-damping, temperature-tunable spin-wave optical elements is provided by superconductors \cite{borst_observation_2023,golovchanskiy_magnetization_2020,ghirri_interplay_2024,yu_efficient_2022}. Superconductors have zero DC electrical resistance, precluding Ohmic dissipation. The strong diamagnetism of superconducting metals placed on top of a magnetic film screens the magnetic stray fields generated by spin waves propagating in the film, modifying the spin wave dispersion \cite{borst_observation_2023,golovchanskiy_magnetization_2020,ghirri_interplay_2024,yu_efficient_2022,zhou_giant_2024}. As the screening strength is tunable via the temperature-dependent London penetration depth, this superconductor-induced modification of the spin-wave dispersion provides opportunities for realizing temperature-tunable spin-wave optics. 

\begin{figure*}[!ht]
\includegraphics[scale=0.97]{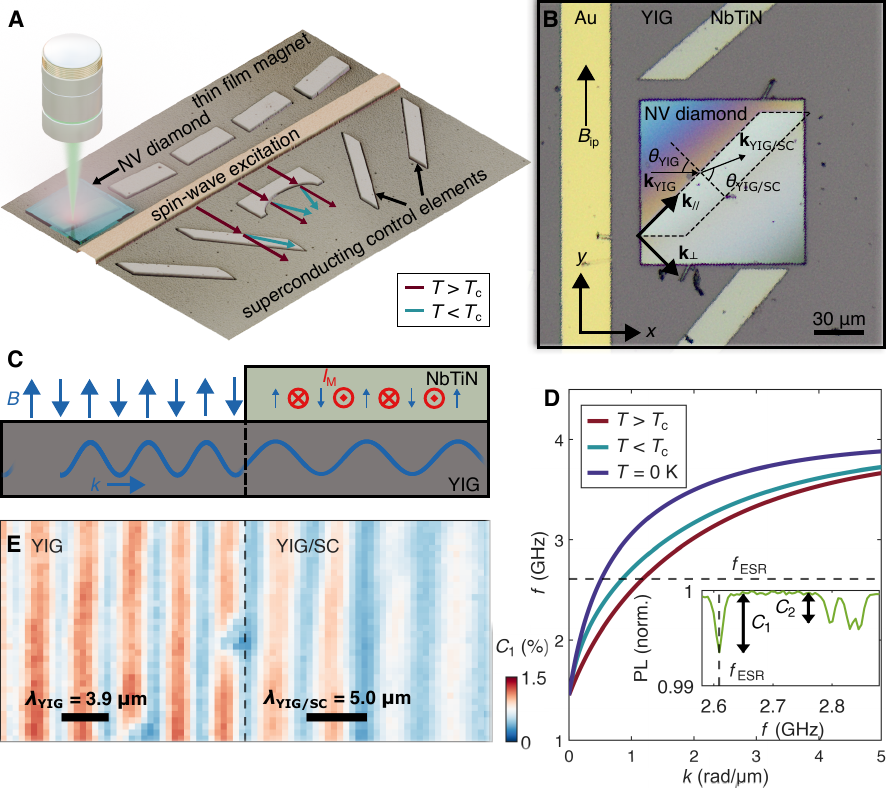}
\caption{\label{fig:spinwave_sc}\textbf{Magnetic imaging of spin waves interacting with superconducting control elements.} (\textbf{A}) Schematic of the experiment. Spin waves in a 0.20 $\upmu$m-thick YIG magnetic film interact with 0.15-$\upmu$m-thick NbTiN islands. A 0.15 $\upmu$m-thick gold microstrip excites the spin waves. A nitrogen-vacancy (NV) diamond enables spin-wave imaging. $T_c$: superconducting critical temperature. (\textbf{B}) Optical micrograph of a $\sim$5-$\upmu$m-thick diamond micromembrane containing NV spins implanted 70\,nm below its bottom surface, placed onto a NbTiN island at 45 deg. w.r.t. the microstrip. The incoming (refracted) spin wave has a wave vector $\mathbf{k}_{\mathrm{YIG}}$ ($\mathbf{k}_{\mathrm{YIG/SC}}$) making an angle $\theta_{\mathrm{YIG}}$ ($\theta_{\mathrm{YIG/SC}}$) w.r.t. the interface normal. The applied magnetic field has an in-plane component $B_{\mathrm{ip}}$ oriented along the microstrip (Damon-Eshbach configuration) and a 35$^\circ$ angle w.r.t. the sample plane to align with one of the four NV orientations in the diamond. (\textbf{C}) Schematic illustration of superconducting Meissner currents ($I_M$) modifying the spin-wave length by screening the magnetic stray fields (blue vertical arrows). (\textbf{D}) Calculated Damon-Eshbach dispersion (spin-wave frequency $f$ vs wavenumber $k$) for YIG covered by NbTiN in the normal ($T>T_c$) and superconducting ($T<T_c$) state. We excite/detect spin waves resonant with the NV ESR frequency $f_{\mathrm{ESR}}$. Inset: Example NV ESR spectrum (photoluminescence (PL) vs microwave drive frequency $f$) with labels $C_{1,2}$ for the ESR contrast of two different NV orientations. (\textbf{E}) Spatial map of the NV ESR contrast, showing spin waves that change in wavelength upon normal traversal of an interface (dashed line) between bare YIG and a YIG-NbTiN (YIG/SC) region. $T_{\mathrm{set}}=6~\mathrm{K}$, $B_{\mathrm{ip}}=5~\mathrm{mT}$, $f_{\mathrm{ESR}}=2.74~\mathrm{GHz}$.}
\end{figure*}

Here, we realize, image, and control the refraction of spin waves using superconducting control elements placed on top of a thin-film magnetic insulator (Fig.~\ref{fig:spinwave_sc}A-B). We characterize the refraction over a wide range of temperatures, frequencies, and incident spin-wave angles, realizing both positive and negative spin-wave refraction and revealing a rich interplay between refracted and secondary spin waves resulting from screening currents in the superconductor. We analyze the observed refraction patterns by combining phase continuity at interfaces with calculations of the superconductor-modified spin wave dispersion.

\section{Experimental Platform and Device Geometry}
Our sample consists of a thin film of yttrium iron garnet (YIG) - a ferrimagnetic insulator with record-low spin-wave damping \cite{yu_magnetic_2014} – with superconducting niobium titanium nitride (NbTiN) elements fabricated on top (Fig.~\ref{fig:spinwave_sc}A-B). In all presented experiments, we apply an external magnetic field to magnetize the YIG along the microstrip used for spin-wave excitation (Damon-Eshbach configuration, Fig.~\ref{fig:spinwave_sc}B). By fabricating the NbTiN elements at various angles relative to the microstrip (Fig.~\ref{fig:spinwave_sc}A), we study the angular dependence of the spin-wave refraction at the interfaces between the bare-YIG and the superconductor-covered YIG regions (labeled as YIG and YIG/SC).

\section{Spin-wave magnetometry with NV centers}

To image the spin waves, we use magnetic imaging based on nitrogen-vacancy (NV) centers in diamond micromembranes placed on top of the sample (Fig.~\ref{fig:spinwave_sc}A-B) \cite{casola_probing_2018}. NV centers have an electron spin-1 that polarizes under green laser excitation and an electronic level structure that yields a spin-dependent photoluminescence. A magnetic field resonant with the NV electron spin resonance (ESR) frequency $f_{\mathrm{ESR}}$ drives spin transitions that are detectable through a reduced photoluminescence \cite{rondin_magnetometry_2014}, while a static magnetic field shifts the ESR frequency (inset Fig.~\ref{fig:spinwave_sc}D). Due to their broad temperature operability, robust diamond host, and ability to image static and dynamic magnetic fields with high spatial resolution, NV magnetometry finds wide applicability in fields ranging from biochemistry, biophysics, geoscience, and condensed matter physics \cite{casola_probing_2018,rondin_magnetometry_2014,schirhagl_nitrogen-vacancy_2014,glenn_micrometer-scale_2017,zhou_quantum_2023}.

NV-imaging of spin waves with few-micrometer wavelengths requires a small, $\lesssim 1~\upmu\mathrm{m}$ distance between the magnetic film and the NV spins due to the exponential decay of the spin-wave stray fields \cite{simon_filtering_2022, abrahams_integrated_2021, zhou_magnon_2021}. To achieve this small distance, we fabricate diamond micromembranes containing a thin layer of NV centers implanted at $\sim 70~\mathrm{nm}$ from their bottom surface \cite{asif_diamond_2024, ghiasi_nitrogen-vacancy_2023} (Supplement). Using a needle, we place these membranes directly on top of target regions of the sample (Fig.~\ref{fig:spinwave_sc}B). The 100-$\upmu\mathrm{m}$ lateral size of the membranes limits the chance of capturing spurious particles (e.g.\ dust) between diamond and sample that would increase the diamond--sample distance. 

We generate spin waves by sending an electrical current with frequency $f_{\mathrm{ESR}}$ through a gold microstrip on the YIG (Fig.~\ref{fig:spinwave_sc}B). This current generates an oscillating magnetic field that excites NV-resonant spin waves in the YIG. These propagating spin waves generate a magnetic stray field that interferes with the direct field of the microstrip, leading to a spatially varying microwave amplitude and corresponding ESR contrast in the NV layer \cite{zhou_magnon_2021,bertelli_imaging_2021,ogawa_wideband_2025}. As such, we can image the spin waves by spatially mapping the NV ESR contrast $C$ (inset Fig.~\ref{fig:spinwave_sc}D). By applying a static magnetic field, we tune $f_{\mathrm{ESR}}$ to a target frequency via the electron Zeeman interaction. In Fig.~\ref{fig:spinwave_sc}E, we show an exemplary image of the change in wavelength of a spin wave as it travels across the interface between a bare-YIG and superconductor-covered YIG region.



\begin{figure}[!ht]
\includegraphics[scale=0.73]{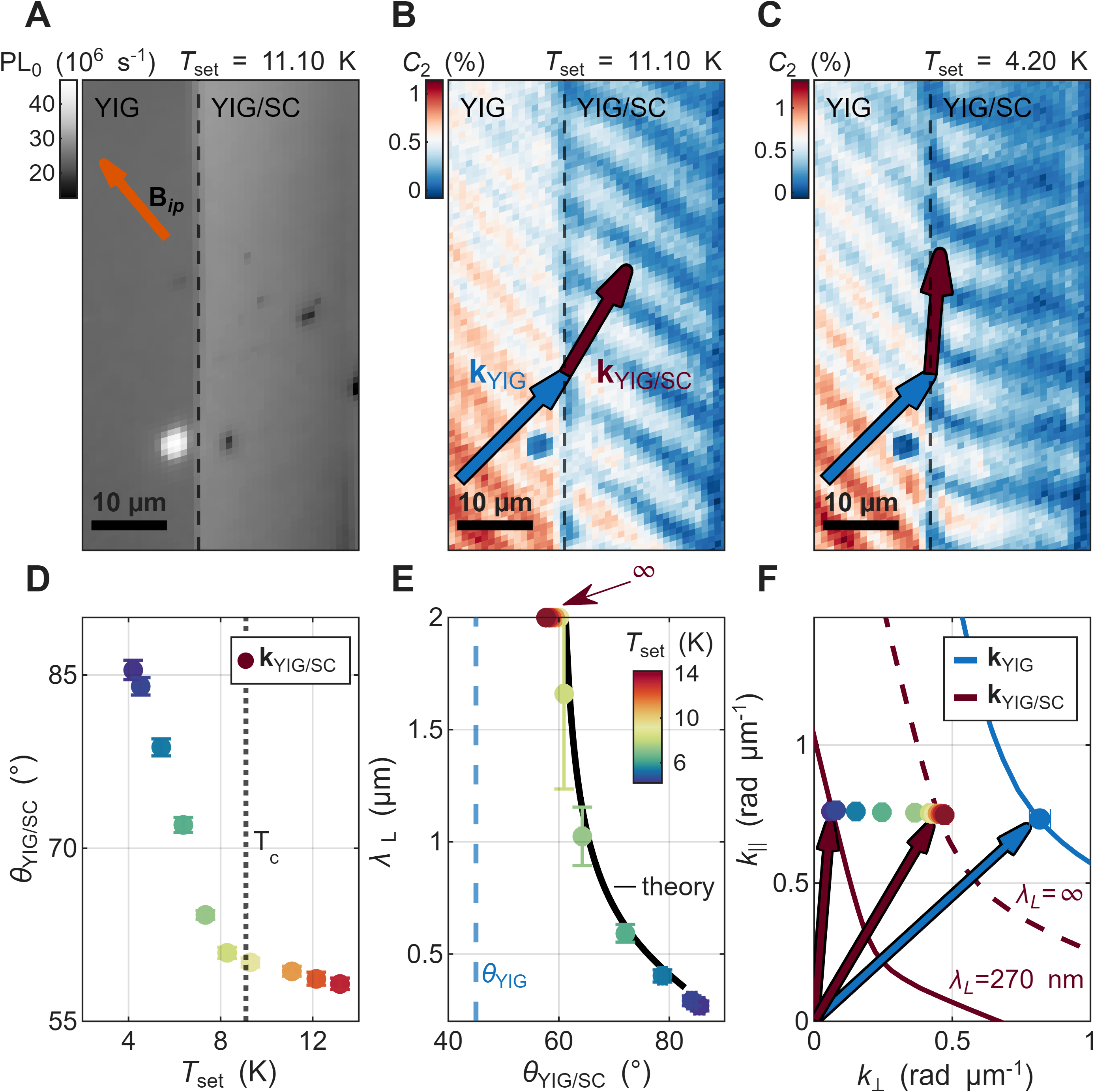}
\caption{\label{fig:temperature_refraction}\textbf{Temperature-controlled, superconductor-induced spin-wave refraction.} (\textbf{A}) Spatial map of NV photoluminescence (PL$_0$) showing the YIG and NbTiN-covered-YIG (YIG/SC) regions of the device in Fig.~\ref{fig:spinwave_sc}B. We rotated the image by 45 deg.\ with respect to Fig.~\ref{fig:spinwave_sc}B to orient the interface vertically. (\textbf{B, C}) Spatial maps of NV ESR contrast with the NbTiN in the normal (\textbf{B}) and superconducting (\textbf{C}) state. A blue (red) arrow indicates the incoming (refracted) wave vector extracted via Fourier analysis (fig.~S1). $B_{\mathrm{ip}} = 13~\mathrm{mT}$, $f_{\mathrm{ESR}} = 2.75~\mathrm{GHz}$. (\textbf{D}) Angle of the refracted spin wave $\theta_{\mathrm{YIG/SC}}$ versus setpoint temperature $T_{\mathrm{set}}$. Dashed line: independently measured NbTiN critical temperature $T_c$ (fig.~S4). (\textbf{E}) London penetration depth ($\lambda_L$), extracted from the observed $\mathbf{k}_{\mathrm{YIG/SC}}$, vs $\theta_{\mathrm{YIG/SC}}$ (fig.~S2). The color-coded temperature enables comparing (\textbf{D--F}). The small refraction above $T_c$ (see also (\textbf{B}) and (\textbf{D})) is attributed to strain- or damage-related anisotropy induced by the NbTiN deposition. (\textbf{F}) Wavevectors of incident and refracted waves decomposed into components parallel $k_{\parallel}$ and perpendicular $k_{\perp}$ to the interface, highlighting the conservation of $k_{\parallel}$. The isofrequency contours of the spin-wave dispersion in the YIG region (blue line) and in the YIG/SC region at the highest (red dashed line) and lowest temperature (red solid line) are overlaid.}
\end{figure}

\section{Temperature-tunable spin-wave refraction}

First, we demonstrate temperature-tunable spin-wave refraction at a YIG-YIG/SC interface. We focus on the device shown in Fig.~\ref{fig:spinwave_sc}B, which has interfaces oriented at 45 deg.\ relative to the microstrip. We apply an external magnetic field along the microstrip, excite NV-resonant Damon--Eshbach spin waves, and image the superconductor-induced change in spin-wave vector by mapping the NV ESR contrast above and below the superconducting transition temperature $T_c$ (Fig.~\ref{fig:temperature_refraction}A--C). Extracting the wave vectors from spin-wave maps taken at different temperatures using Fourier transformation (fig.~S1), we observe an increasingly strong, superconductor-induced change in the direction of $\mathbf{k}_{\mathrm{YIG/SC}}$ as we lower the temperature towards the base temperature of our cryostat (Fig.~\ref{fig:temperature_refraction}D). In addition, we observe a small, temperature-dependent change in $\mathbf{k}_{\mathrm{YIG/SC}}$ above $T_c$, presumably due to a change in the YIG magnetic anisotropy induced by the NbTiN deposition (supplementary text, section~S4) resulting in strain or damage of the underlying YIG.

We analyze the temperature dependence of the observed refraction using a model of the superconductor-induced change in the spin-wave dispersion \cite{borst_observation_2023,yu_efficient_2022}. In this model, the diamagnetic response of the superconductor screens the spin-wave magnetic stray fields (Fig.~\ref{fig:spinwave_sc}C), modifying the dispersion of dipolar spin waves (Fig.~\ref{fig:spinwave_sc}D). As such, the spin-wave dispersion depends on the London penetration depth, enabling us to determine the penetration depth at each temperature from the observed wave vector $\mathbf{k}_{\mathrm{YIG/SC}}$ by fitting (Fig.~\ref{fig:temperature_refraction}E). We observe the penetration depth decreasing with decreasing temperature (indicated by the color coding in Fig.~\ref{fig:temperature_refraction}E), reaching $0.27~\upmu\mathrm{m}$ at the base temperature of our cryostat. This is in the range of values reported for sputtered NbTiN thin films, varying from $0.20~\upmu\mathrm{m}$ to $0.38~\upmu\mathrm{m}$ depending on the deposition conditions and film quality \cite{lee_penetration_2024,yu_fabrication_2005}. The tunability of the superconductor-induced spin-wave refraction via the London penetration depth highlights the potential for realizing temperature-tunable spin-wave optics. 

Both before and after refraction, the wave vectors $\mathbf{k}_{\mathrm{YIG}}$ and $\mathbf{k}_{\mathrm{YIG/SC}}$ must lie on the isofrequency contours of the respective regions (Fig.~\ref{fig:temperature_refraction}F). Decomposing the wave vectors into components parallel and perpendicular to the interface (Fig.~\ref{fig:temperature_refraction}F) shows that the parallel component is conserved, $ |\mathbf{k}_{\mathrm{YIG}}|\sin\theta_{\mathrm{YIG}} = |\mathbf{k}_{\mathrm{YIG/SC}}|\sin\theta_{\mathrm{YIG/SC}}. $ As such, our model of the spin-wave dispersion in the YIG/SC region combined with the conservation of the parallel component of the wave vector enables predicting the refracted wave vector for any incident wave vector and penetration depth. The black line Fig.~\ref{fig:temperature_refraction}E shows the predicted refraction for the incident wave-vector angle of 45 deg.

\begin{figure} [!ht]
\includegraphics[scale=0.73]{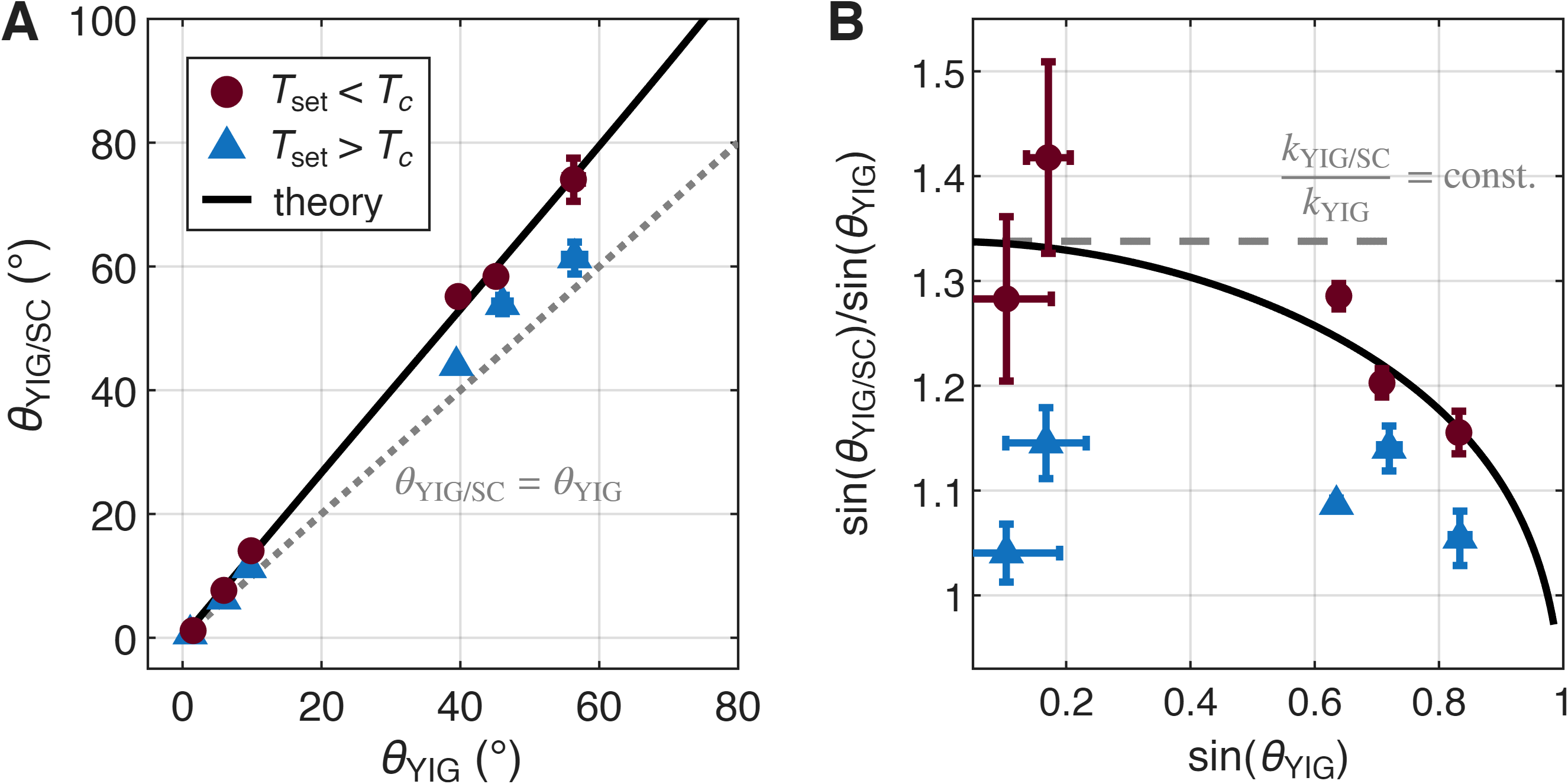}
\caption{\label{fig:snell}\textbf{Refractive index and Snell's law.} (\textbf{A}) Angle $\theta_{\mathrm{YIG/SC}}$ of the refracted wave vector vs angle of the incident wave vector $\theta_{\mathrm{YIG}}$, measured above $T_c$ (blue dots) and below (red dots). The waves break away from the interface normal ($\theta_{\mathrm{YIG/SC}}>\theta_{\mathrm{YIG}}$). The dashed line indicates $\theta_{\mathrm{YIG}}=\theta_{\mathrm{YIG/SC}}$. Each incident angle corresponds to a superconducting island oriented at that angle w.r.t.\ the microstrip (Fig.~\ref{fig:spinwave_sc}A). Angles extracted by Fourier transformation of spatial spin-wave maps as in Fig.~\ref{fig:temperature_refraction}. See figures S10--15 for all spatial maps. $B_{\mathrm{ip}}=8.0~\mathrm{mT}$, $f_{\mathrm{ESR}}=2.603~\mathrm{GHz}$. As in Fig.~\ref{fig:temperature_refraction}A, we observe a small refraction above $T_c$ attributed to a NbTiN-deposition-induced anisotropy. (\textbf{B}) Ratio of the sines of the angles of the incident and refracted wave vectors (same data as in A). The observed change of the ratio indicates a change of the effective refractive index experienced by the spin waves in the YIG/SC region. Black line: calculated refraction for $\lambda_{\mathrm{L}}=0.8~\upmu\mathrm{m}$, and anisotropy parameters $B_u^{\mathrm{YIG/SC}}=10~\mathrm{mT}$ and $B_u^{\mathrm{YIG}}=1.5~\mathrm{mT}$ obtained from an independent measurement (fig.~S3). Error bars are $\pm 1$ s.d.}
\end{figure}

To characterize the effective refractive index experienced by the spin waves in the YIG/SC region, we measure the angle of the refracted wave vector over a range of incident wave-vector angles (Fig.~\ref{fig:snell}A). We use the same Damon--Eshbach wave for all angles of incidence, enabled by using superconducting elements with different orientations with respect to the microstrip. We image the refraction by placing NV-diamond micromembranes onto each element (fig.~S21), orienting the membranes in the same way such that we can apply the same drive frequency and magnetic field.

We observe that the refracted wave breaks away from the interface normal ($\theta_{\mathrm{YIG/SC}} > \theta_{\mathrm{YIG}}$, Fig.~\ref{fig:snell}A) and that $\theta_{\mathrm{YIG/SC}}$ increases linearly with $\theta_{\mathrm{YIG}}$. This linear increase indicates that the effective refractive index of the YIG/SC region changes. Using Snell's law, $ n_{\mathrm{YIG}}\sin\theta_{\mathrm{YIG}} = n_{\mathrm{YIG/SC}}\sin\theta_{\mathrm{YIG/SC}}$ with $n_{\mathrm{YIG}}$ ($n_{\mathrm{YIG/SC}}$) the refractive index of the YIG (YIG/SC) region, we highlight this change by plotting the ratio $ \frac{\sin\theta_{\mathrm{YIG/SC}}}{\sin\theta_{\mathrm{YIG}}}$ as a function of the angle of the incident wave vector (Fig.~\ref{fig:snell}B). We observe that the effective refractive index decreases with increasing wave vector angle $\theta_{\mathrm{YIG/SC}}$. This decrease is captured by our geometric analysis of the refraction, similar to that shown in Fig. 2F (black line Fig. 3B).


\section{Positive and negative phase refraction}

The dispersion of spin waves in magnetic films with in-plane magnetization is characterized by strongly anisotropic, hyperbolic isofrequency contours (see e.g.\ Fig.~\ref{fig:temperature_refraction}F and fig.~S2) \cite{kalinikos_theory_1986} that derive from the anisotropic dipolar fields generated by the spins in the film. Such hyperbolic dispersions lead to large angles between the phase and group velocities and are of interest in optics \cite{poddubny_hyperbolic_2013} as they enable phenomena such as negative refraction, hyper-focusing and ways to control wave transport that are absent in systems with isotropic dispersions \cite{smith_electromagnetic_2003}.

Here we demonstrate that our superconducting control elements enable tuning the spin-wave refraction from positive to negative. We focus on the same device as in Fig.~\ref{fig:temperature_refraction}/Fig.~\ref{fig:spinwave_sc}B and change the wavelength of the impingent spin wave by changing the excitation frequency. To retain resonance with the NV ESR frequency, we simultaneously change the magnetic field. For large frequency/small impingent wavelength (Fig.~4A), the refracted spin-wave vector breaks away from the interface normal. Upon decreasing the frequency (Fig.~4B), the change in wave vector increases, causing the refracted wave fronts to be oriented nearly perpendicular to the interface. For isotropic dispersions, this point would mark the onset of total internal reflection, where the incident wave becomes fully reflected and only has an evanescent tail beyond the interface. Here, when tuning to even smaller frequencies (Fig.~4C), we instead observe a refracted wavevector of which the component perpendicular to the interface has become negative.

\begin{figure}[!ht]
\includegraphics[scale=0.73]{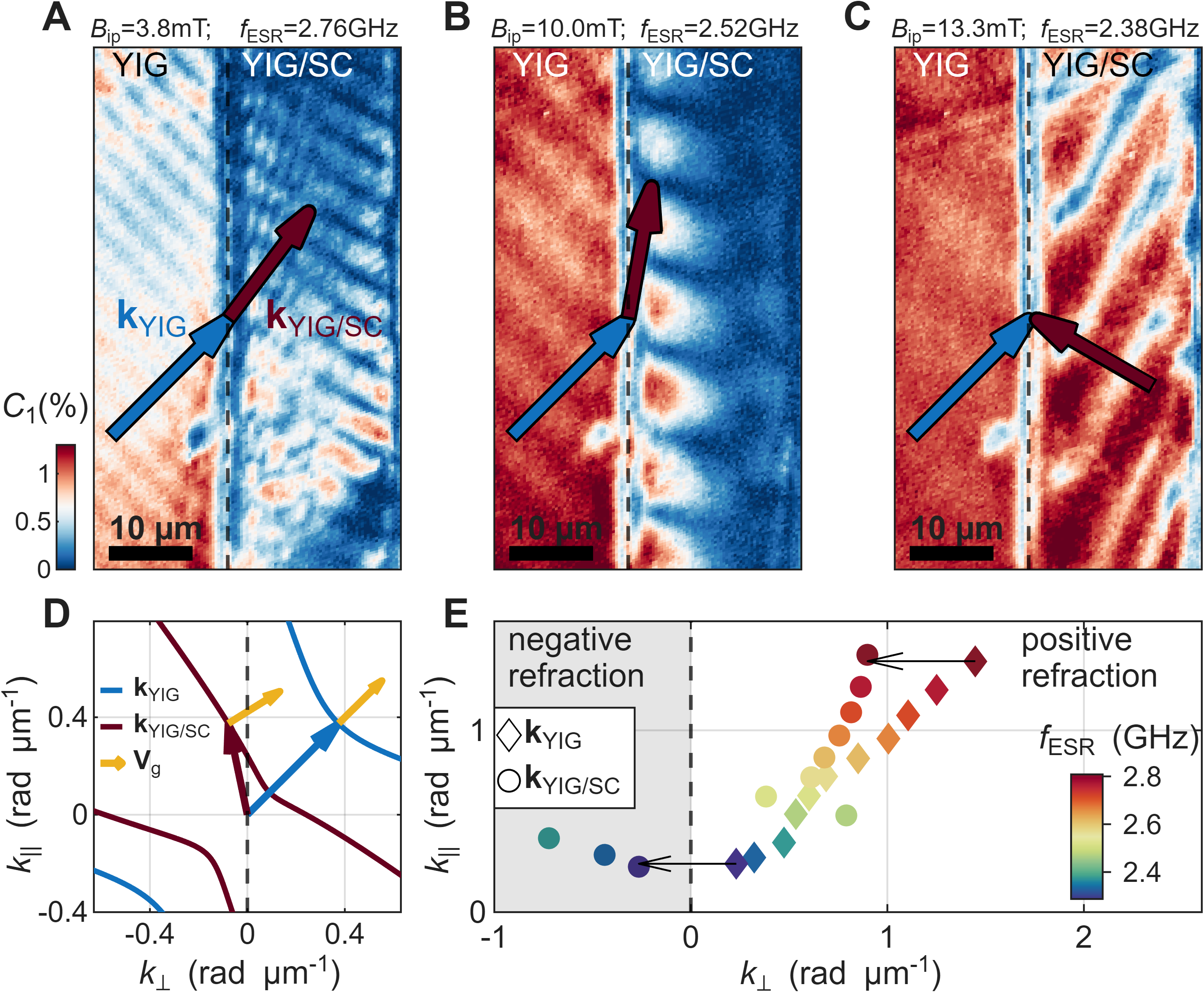}
\caption{\label{fig:negative_refraction}\textbf{Positive-to-negative refraction by a superconducting island.} (\textbf{A--C}) Spatial maps of the NV ESR contrast at different drive frequencies. Reducing the frequency (from \textbf{A} to \textbf{C}) causes stronger refraction and the perpendicular component of the wave vector to change sign in \textbf{C}. To retain resonance with the NV frequency, we simultaneously tune the applied magnetic field. $T_{\mathrm{set}}=4.4~\mathrm{K}$. (\textbf{D}) Geometric analysis of the refraction in $k$-space. Both before and after refraction, the wave vector must lie on the isofrequency contour of the spin-wave dispersion (blue line for bare YIG and red line for YIG/SC, calculated for the field and frequency of (\textbf{C}) and $\lambda_{\mathrm{L}}=0.4~\upmu\mathrm{m}$). The conservation of the wave-vector component parallel to the interface (dashed line) yields the change in perpendicular component. In the case shown, the perpendicular component of the wave vector becomes negative, but that of the group velocity $\mathbf{v}_g$ (yellow arrows) does not. (\textbf{E}) Incident ($\mathbf{k}_{\mathrm{YIG}}$) and refracted ($\mathbf{k}_{\mathrm{YIG/SC}}$) wave vectors for a range of different frequencies $f_{\mathrm{ESR}}$ (color-coded), showing the conservation of the parallel component and the transition to the negative refraction regime. The black arrow indicates the change in wave vector upon refraction.}
\end{figure}

The change from positive to negative refraction can be understood geometrically from the hyperbolic isofrequency contours of the spin-wave dispersion (Fig.~\ref{fig:negative_refraction}D) as the wave vectors $\mathbf{k}_{\mathrm{YIG}}$ and $\mathbf{k}_{\mathrm{YIG/SC}}$ must lie on these contours. For the field and frequency of Fig.~\ref{fig:negative_refraction}C, this geometric analysis reproduces the observed negative refraction of the spin-wave vector (Fig.~\ref{fig:negative_refraction}D). The analysis also shows that the group velocity, given by the gradient of the spin-wave dispersion, remains positive with respect to the interface, consistent with the strong spin-wave signal still observed in Fig.~\ref{fig:negative_refraction}C underneath the superconductor.

To provide an overview of the transition from positive to negative refraction, we measure the incident and refracted wave vectors over a range of spin-wave frequencies (Fig.~\ref{fig:negative_refraction}E, fig.~S8). For all frequencies, we observe that the component of the wave vector parallel to the interface is conserved, whereas the perpendicular component changes from positive to negative as we lower the frequency of the excited spin wave. These results highlight that the tunability of superconductors enables realizing negative refraction or other non-conventional spin-wave optical effects.

\section{Spin wave transmission through superconducting elements}

\begin{figure*}[!ht]
\includegraphics[scale=1.]{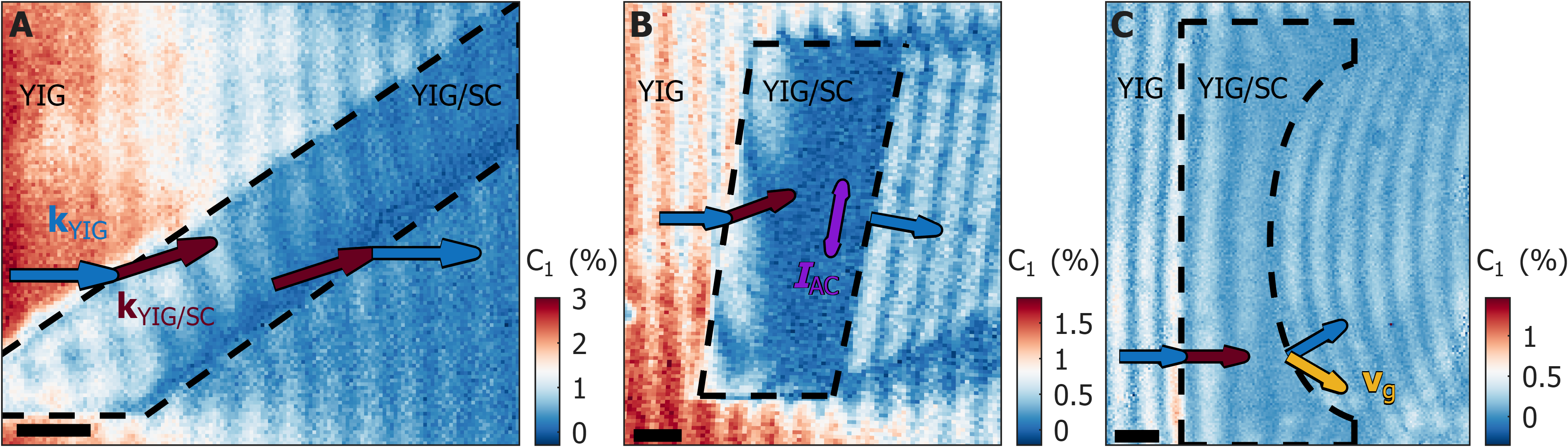}
\caption{\label{fig:steering}\textbf{Refraction and secondary emission of spin waves by superconducting elements.} (\textbf{A}) Spatial maps of the NV ESR contrast showing a spin wave interacting with a superconducting element oriented at 55 deg.\ w.r.t.\ to the incoming wave vector $\mathbf{k}_{\mathrm{YIG}}$. The spin wave is excited by the gold microstrip (left outside the image), refracts at the 1st interface, and returns to its original direction upon refraction at the 2nd interface. $B_{\mathrm{ip}}=8.0~\mathrm{mT}$, $f_{\mathrm{ESR}}=2.60~\mathrm{GHz}$, $T_{\mathrm{set}}=4.6~\mathrm{K}$. (\textbf{B}) Secondary spin-wave emission by a superconducting element oriented at 10 deg.\ with respect to the incoming $\mathbf{k}_{\mathrm{YIG}}$. The wave fronts to the right of the element run parallel to the edge of the element, indicating they are excited by a current running parallel to the edge of the element. $B_{\mathrm{ip}}=10~\mathrm{mT}$, $f_{\mathrm{ESR}}=2.52~\mathrm{GHz}$, $T_{\mathrm{set}}=4.2~\mathrm{K}$. (\textbf{C}) Radially converging spin-wave fronts excited by a superconducting element with a curved interface. Yellow arrow: local group velocity direction indicating a diverging nature of the intensity. $B_{\mathrm{ip}}=10~\mathrm{mT}$, $f_{\mathrm{ESR}}=2.57~\mathrm{GHz}$, $T_{\mathrm{set}}=5.9~\mathrm{K}$. All scale bars $10~\upmu\mathrm{m}$.}
\end{figure*}

To further study the potential of superconductors for spin-wave optics, we now image the spin-wave patterns that emerge beyond the 2nd interface of the superconducting elements. We first image a spin wave impingent from the left onto an element oriented at 55 deg.\ with respect to the incoming wave vector (Fig.~\ref{fig:steering}A). We observe that the refraction at the 1st interface causes the wave vector to change direction underneath the superconductor (as in Fig.~\ref{fig:temperature_refraction}), and to return to its original direction as expected for refraction at the 2nd interface. This observation shows that superconducting elements can be used as spin-wave phase shifters or beam translators.

Next, we study a superconducting element oriented at 10 deg.\ with respect to the incoming wave vector (Fig.~\ref{fig:steering}B). Contrary to the expectation for spin-wave refraction, here we find that the spin-wave fronts emerging beyond the second interface are oriented parallel to the interface. We attribute this effect to a screening current in the superconductor that runs parallel to the edges of the element induced by the direct magnetic field of the microstrip. This current excites a secondary spin-wave pattern with wave fronts parallel to the element's edge. 

We find these secondary spin waves to dominate beyond the 2nd interface for superconducting elements that make a small angle with the incoming wave (Fig.~\ref{fig:steering}B) and the refraction to dominate for elements that make a large angle (Fig.~\ref{fig:steering}A, fig.~S15). To understand the relative strength of the refracted and secondary spin-wave patterns, we first note that, the larger the angle of the superconducting element, the smaller the wavelength of the secondary spin waves with wavefronts parallel to its edge. Because both the excitation and the NV detection efficiency decrease with decreasing spin-wave length, the secondary patterns become less prominent for large angle devices.

To further develop spin-wave steering using superconducting elements, we fabricated a superconducting island with a lens-shaped interface (Fig.~\ref{fig:steering}C). Above $T_c$, the spin waves are largely unaffected by the metal (fig.~S20) apart from small scattering effects at the edges. In contrast, below $T_c$, the superconducting element causes a wavelength change underneath the element and a radially shaped, converging spin-wave pattern beyond it (Fig.~\ref{fig:steering}C). While the converging spin-wave fronts resemble the focusing of the spin wave by a positive lens, the anisotropic dispersion implies that the corresponding group velocity distribution is diverging (Fig.~\ref{fig:steering}C). As such, the element acts as an optical element that diverges the spin-wave intensity.

\section{Conclusions}
We demonstrated superconductor-controlled refraction and steering of dipolar spin waves in a magnetic thin film. By varying temperature, frequency, and incidence angle, we characterized a broad range of spin-wave refraction angles and tuned into regimes of positive and negative refraction. The angular dependence of the refracted spin waves was accurately captured by a geometrical analysis of the wave vectors and calculated isofrequency contours in the bare and superconductor-covered YIG regions.

Studying the spin-wave patterns emerging beyond the superconducting elements revealed an interplay between refracted spin waves and secondary spin-wave patterns attributed to direct inductive coupling between the microstrip and the superconductor. We anticipate that this secondary pathway could be enhanced or suppressed relative to the refracted patterns by tuning the size and shape of the superconducting island to realize or avoid resonance, or by changing the excitation microstrip into a balanced coplanar waveguide that generates less far field.

A key advantage of superconductor-based spin-wave refraction is its temperature-tunability, which may be enhanced beyond the presented results by using a superconductor with smaller penetration depth. The spin-wave propagation through regions with tunable refractive-index resembles that of light traveling through tunable dielectrics. As such, the demonstrated superconductor-based spin-wave control paves the way towards spin-wave optical setups that include superconducting lenses, beam splitters and filters for realizing microwave control at the micrometer-scale.

\section*{Author contributions}
\textbf{PV}, \textbf{MiB}, and \textbf{TS} conceived the project. \textbf{PV}, \textbf{MAB}, \textbf{MiB}, \textbf{TO}, and \textbf{TS} developed the methodology. \textbf{PV}, \textbf{MAB}, and \textbf{MiB} carried out the investigation. \textbf{PV}, \textbf{MAB}, and \textbf{MiB} performed the data visualization. \textbf{TS} and \textbf{RD} acquired funding for the project. \textbf{TS} administered the project. \textbf{RD} and \textbf{TS} supervised the research. \textbf{PV} and \textbf{TS} wrote the original manuscript draft. \textbf{PV}, \textbf{MAB}, and \textbf{TS} reviewed and edited the manuscript, with input from all co-authors.

\begin{acknowledgments}
We thank I.O. Robertson for commenting on the manuscript. This project has received funding from the European Research Council (ERC) under the 5 European Union’s Horizon 2020 research and innovation programme (grant agreement No 101170480 MAGICWAVE) and from by the Dutch Research Council (NWO) under awards VI.Vidi.193.077, NGF.1582.22.018, and OCENW.XL21.XL21.058.
\end{acknowledgments}

\clearpage
\bibliography{Refractionpaperv2.bib}

\end{document}


\beginsupplement 

\begin{center}
\textbf{{\huge Supplementary material for}}\break\break

\textbf{{\Large Temperature-tunable spin-wave refraction using superconducting control elements }}\break\break

{\large P. H. Vree}$^{1}$, {\large M.A. Bouma}$^{1}$, {\large M. Borst}$^{1}$, {\large T. T. Osterholt}$^{2}$, {\large R. A. Duine}$^{2,3}$, {\large T. van der Sar}$^{1}$

\vspace{0.5cm} \small{ $^1$Department of Quantum Nanoscience, Kavli Institute of Nanoscience, Delft University of Technology, 2628 CJ Delft, The Netherlands

$^2$Institute for Theoretical Physics, Utrecht University, 3584 CC Utrecht, The Netherlands

$^3$Department of Applied Physics, Eindhoven University of Technology, 5600 MB Eindhoven, The Netherlands } \vspace{0.3cm} 

Corresponding author: \texttt{T.vanderSar@tudelft.nl}

\end{center}
\tableofcontents

\section{Materials and Methods}
\subsubsection{Sample fabrication}
The (111)-oriented, $0.20~\upmu$m-thick yttrium iron garnet (YIG) film used in this work was grown on a $500~\upmu$m-thick gadolinium gallium garnet (GGG) substrate by Matesy GmbH. Using e-beam lithography and metal evaporation, we deposited a $150$-nm-thick gold (Au) microstrip onto the YIG film using a $5$-nm-thick titanium (Ti) adhesion layer. Using e-beam lithography and sputtering, we fabricated the superconducting elements from $150$-nm-thick niobium titanium nitride (NbTiN).

\subsubsection{Diamond membrane fabrication and placement}
We start by having a $4\times4\times0.5~\mathrm{mm}^3$ electronic-grade diamond (grown by Element 6 Inc.) cut into $2\times2\times0.05~\mathrm{mm}^3$ platelets by Almax Easylabs. After nitrogen implantation at $54~\mathrm{keV}$ and a density of $10^5$ ions/$\upmu\mathrm{m}^2$ by CuttingEdgeIons, we anneal the diamond at $800^\circ$C to create NV centers. This yields an estimated NV density of $10^3~\upmu\mathrm{m}^{-2}$ at approximately $70~\mathrm{nm}$ below the surface~\cite{pezzagna_creation_2010}. 

Next, we etch $100\times100\times5~\upmu\mathrm{m}^3$ and $100\times250\times5~\upmu\mathrm{m}^3$ micromembranes into an NV-diamond platelet. We deposit a $500$-nm-thick SiN hard mask via plasma-enhanced chemical vapor deposition (PECVD), apply a CSAR-13 resist, and write the membrane array into the resist using e-beam lithography (Raith EBPG5200+). We transfer the pattern into the SiN using CHF$_3$/O$_2$ reactive ion etching (RIE) and subsequently into the diamond by $7~\upmu$m-deep O$_2$ RIE. Then we flip the diamond and O$_2$-etch using a quartz wafer with a $1.2\times1.2~\mathrm{mm}^2$ opening as a hard mask to create membranes that are only attached to the host diamond by two or four $1~\upmu$m-wide tethers. In the final step we remove the SiN layer with a 10-minute HF acid dip and clean the diamond for 10 minutes in HNO$_3$. 

To position a diamond membrane onto the YIG sample, we glue the side of the host diamond to a stainless-steel needle and bring it close to the target location using a micromanipulator. We then press a micromembrane onto the sample using a needle in a second micromanipulator, breaking the holding bars. Using the same needle, we can move the membrane to optimize its position if needed. Most membranes do not displace during subsequent maneuvering, wire bonding, or vertical mounting in the cryostat. Some membranes did fall off during cooldown, likely due to sample contamination resulting in increased stand-off distance and reduced van der Waals forces.


\subsubsection{Experimental setup}
We measure the NV photoluminescence using a home-built confocal microscope and a closed-cycle cryostat (Montana Cryostation S100) with optical access and a room-temperature microscope objective (NA = 0.85). We position the sample using slip-stick positioners (Attocube ANPx101, ANPz102) and scan a $520$ nm continuous-wave laser (Coherent Obis 520LX) over our sample using a fast-steering mirror (Newport FSM300). Using two avalanche photodiodes (Excelitas SPCM-AQRH-13) or a Si-PIN photodiode (Femto PWPR-2K-SI), we detect the NV photoluminescence after 600 nm long-pass filtering. We excite spin waves and drive the NV centers by sending a current generated by a microwave generator (Rohde \& Schwarz SGS100A) through the Au microstrip. 

To tune the NV electron spin resonance (ESR) frequency and magnetize the YIG, we apply a magnetic field using a cylindrical neodymium magnet (N45, Supermagnete S-35-20-N, 20 mm tall, 35 mm diameter) mounted outside the cryostat on a combined rotational (Zaber T-RS60) and translational (Zaber X-LRT0250AL-E08C) stage. This configuration allows us to generate magnetic fields within the $yz$ plane with magnitudes up to 30 mT.

\section{Supplementary Text}
\subsection{Magnet calibration} 
To relate the stage coordinates to the magnetic field at the sample, we measure the NV electron spin resonance (ESR) spectrum as a function of the stage position $(\theta_s,r_s)$. A typical method for extracting the magnetic field from such spectra is to first determine the ESR frequencies and subsequently fit these frequencies to the NV Hamiltonian. However, extracting the ESR frequencies and assigning them consistently to the same NV orientation is challenging when resonances overlap and cross as the magnet position is varied.

We developed a calibration procedure to overcome this challenge. In this approach, we fit the measured ESR spectra over the full range of stage positions using a model that calculates the ESR spectra directly from the magnet position. The fit parameters include the relative orientations and positions of the diamond and magnet, the center and axis of rotation of the magnet, the direction of the magnet's linear motion, the magnetization of the magnet, and the contrast and linewidths of the NV ESR transitions. This procedure avoids the complications associated with overlapping ESR resonances and enables determination of the magnetic field at the sample for all magnet-stage positions required for the spin-wave dispersion analysis presented below. The uncertainty in the calibrated magnetic field is estimated to be $0.3~\mathrm{mT}$.

\subsection{Extraction of spin-wave wave vectors}
We extract the spin-wave vectors shown in Fig.~2 to Fig.~5 of the main text from the spatial NV ESR contrast maps using a two-dimensional Fourier analysis. Figure~S1 shows all spin-wave maps used in Fig.~2D of the main text together with the extraction of the wave vectors via Fourier analysis. We first rotate the measured maps to align the interface along the vertical direction, interpolate the data onto a regular grid, and separate the map into the bare-YIG and NbTiN-covered YIG regions. To suppress boundary effects, we apply a Hann window and zero-pad the data to increase the resolution in $k$-space.

\begin{figure*}[!ht]
\includegraphics{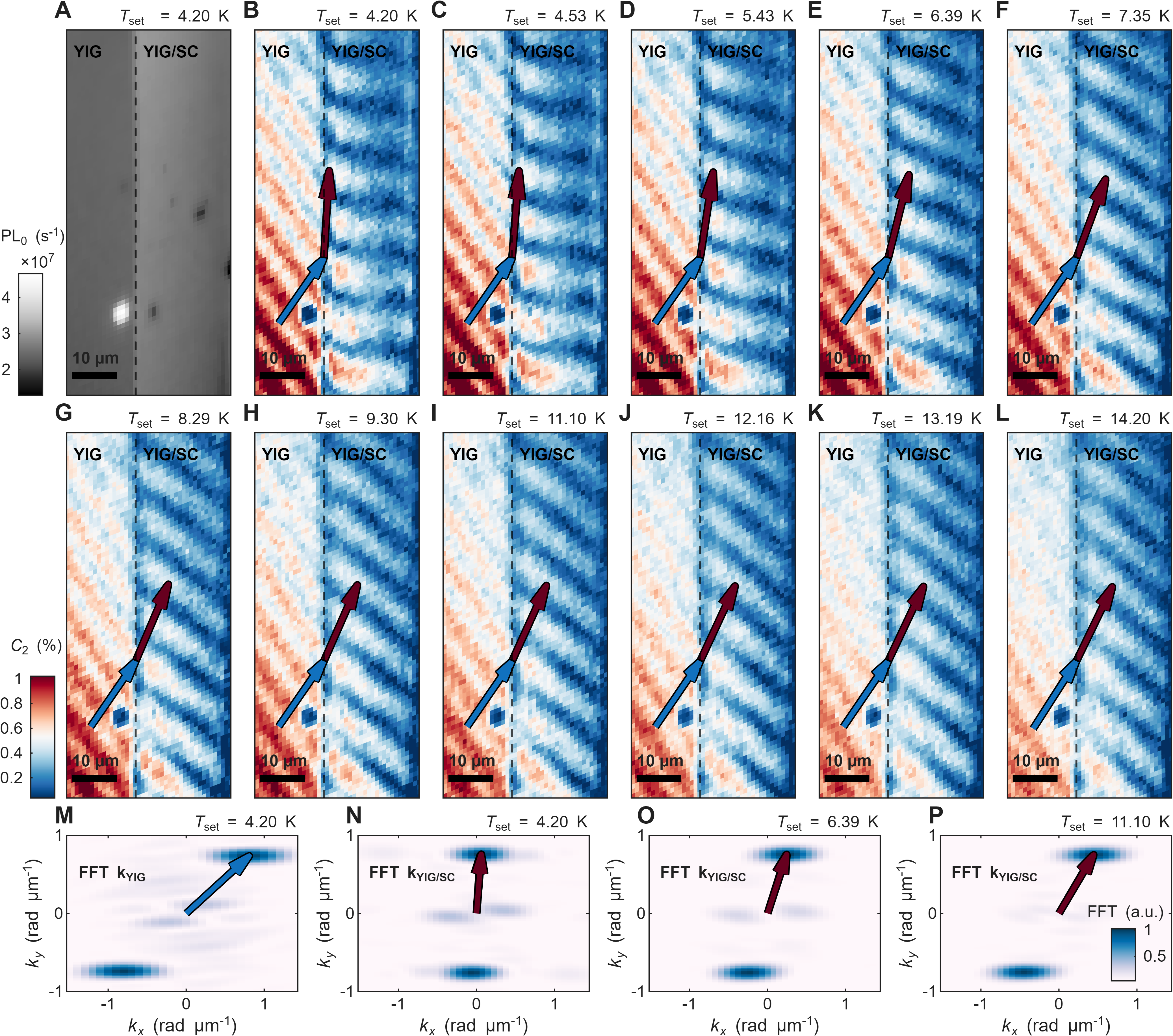}
\caption{ \textbf{Temperature-tunable spin-wave refraction and its Fourier analysis.} (\textbf{A}) Spatial map of the NV photoluminescence, showing the YIG--YIG/SC interface of a NbTiN element oriented at $45^\circ$ with respect to the microstrip (cf. Fig.~2A of the main text). (\textbf{B--L}) Spatial maps of the NV electron spin resonance contrast, $C_2$, for a range of temperatures across the superconducting transition ($T_c = 9.1~\mathrm{K}$). (\textbf{M--P}) Two-dimensional Fourier transforms (FFT) illustrating the extraction of spin-wave vectors in the YIG and YIG/SC regions. Labels $\mathbf{k}_{\mathrm{YIG}}$ and $\mathbf{k}_{\mathrm{YIG/SC}}$ indicate the region used for the analysis, and arrows mark the peak positions extracted via a Gaussian fit. A Hann window and zero-padding are applied prior to the FFT to suppress boundary effects. (a.u.: arbitrary units). } \label{fig:S1}
\end{figure*}

The resulting Fourier spectra exhibit peaks at the wave vectors of the spin-wave modes (Fig.~\ref{fig:S1}M--P). We extract the wave vector from the center of a Gaussian fit to a peak within a selected region of the Fourier spectrum. The uncertainty in the extracted wave vector is estimated from the full width at half maximum (FWHM) of the peak, scaled by the signal-to-noise ratio defined as the ratio between the peak amplitude and the average signal in the Fourier spectrum.

For some spin-wave maps (e.g. Fig.~4A of the main text), we observe multiple spin-wave modes and corresponding peaks in the Fourier spectrum. The origin of these additional modes may be related to spin-wave reflection or scattering. In these cases, we identify the refracted mode by following the gradual evolution of the refraction pattern across the dataset.

\subsection{Spin-wave dispersion in the YIG and YIG/SC regions}

Here we compute the spin-wave dispersion in the YIG and YIG/SC regions. Building on Ref.~\cite{borst_observation_2023}, we extend the description of the system to include (i) uniaxial and cubic magnetic anisotropy, (ii) an equilibrium magnetization that is slightly out of plane, and (iii) the back-action of the superconductor on spin waves in this tilted magnetization.

We found these extensions to be essential for describing the spin-wave dispersion in the bare-YIG and YIG/NbTiN regions. In particular, the refraction observed even above $T_c$ (see e.g.\ Fig.~2D of the main text) is captured by including a change in the uniaxial anisotropy between the two regions, which we attribute to strain or damage caused by the NbTiN deposition. Including the cubic anisotropy of YIG leads to a slightly out-of-plane magnetization at zero field. The out-of-plane angle increases upon applying the magnetic field along one of the NV orientations in the diamond crystal. The spin-wave dispersions derived below enable extraction of the anisotropy parameters and the London penetration depth from the measured wave vectors.

The free energy of our YIG film is given by 
\be F = F_{\mathrm{Z}} + F_{\mathrm{ex}} + F_{\mathrm{dip}} + F_{\mathrm{u}} + F_{\mathrm{cubic}}, \label{eq:free_energy} \ee 
corresponding to the Zeeman, exchange, dipolar, and anisotropy (uniaxial and cubic) contributions, respectively. 
The equilibrium magnetization direction $\mathbf{m}_0$ is obtained by minimizing $F$. 
The spin-wave dispersion is then derived from the dynamics of small deviations around this equilibrium using the Landau--Lifshitz (LL) equation
\be \dot{\mathbf{m}} = -\gamma\, \mathbf{m} \times \left( \mathbf{B}_{\mathrm{eff}} + \mathbf{B}_{\mathrm{SC}} \right), \label{eq:LL} \ee
where $\mathbf{m}$ is the unit magnetization, $\gamma$ is the gyromagnetic ratio, $\mathbf{B}_{\mathrm{eff}}$ is the effective magnetic field derived from the free energy, and $\mathbf{B}_{\mathrm{SC}}$ represents a dynamic field arising from the superconducting back-action.

To compute the spin-wave dispersion from the free energy, we employ the Smit--Beljers formalism~\cite{smit_ferromagnetic_1955}. In this approach, the dispersion is given by
\be \omega = \sqrt{ \left(F_{xx}-F_z\right) \left(F_{yy}-F_z\right) - F_{xy}^{\,2} }, \label{eq:smit_beljers} \ee
where 
\be F_{ij} = \left. \frac{\partial^2 F} {\partial m_i \partial m_j} \right|_{m_x=m_y=0,m_z=1}, \qquad i,j\in\{x,y\}, \label{eq:Fij} \ee and \be F_z = \left. \frac{\partial F} {\partial m_z} \right|_{m_x=m_y=0,m_z=1}. \label{eq:Fz} \ee 
These quantities are derivatives of the free energy in the \textit{magnet frame}, defined as the coordinate system in which the equilibrium magnetization is parallel to the $z$-axis. Throughout this section, we express the free energy in units of angular frequency ($\mathrm{s}^{-1}$). 

We consider a magnetic film in the $x'y'$ plane occupying the region $-t<z<0$, with a superconducting layer on top occupying the region $0<z<h$. The free-energy contributions are defined in the laboratory frame and subsequently transformed into the magnet frame. 

The laboratory frame is denoted by $(\hat{\mathbf{x}}',\hat{\mathbf{y}}',\hat{\mathbf{z}}')$, with unit magnetization 
$\mathbf{m}' = (m_x',m_y',m_z')^{T}, $
which is parameterized by the spherical angles $(\theta_m,\phi_m)$.
The magnet frame is defined as the coordinate system in which $\mathbf{m}_0 = (0,0,1)^T. $ 

The magnet and laboratory frames are related through the rotation 
$ \mathbf{m}' = R(\theta_m,\phi_m)\mathbf{m}, \label{eq:rotation_relation}$ with rotation matrix \be R(\theta,\phi) = \begin{pmatrix} \cos\theta\cos\phi & -\sin\phi & \sin\theta\cos\phi \\ \cos\theta\sin\phi & \cos\phi & \sin\theta\sin\phi \\ -\sin\theta & 0 & \cos\theta \end{pmatrix}. \label{eq:rotation_matrix} \ee 

In the following sections we present the quantities $F_{ij}$ and $F_z$ in the magnet frame for each free-energy contribution.

\textbf{Zeeman energy.}
The applied magnetic field in the laboratory frame is 
\be \mathbf{B} = B_0 \hat{\mathbf{B}}, \qquad \hat{\mathbf{B}} = (\sin\theta_B\cos\phi_B, \sin\theta_B\sin\phi_B, \cos\theta_B). \ee 
The Zeeman energy reads \be F_{\mathrm Z} = -\omega_B \left( \mathbf{m}'\cdot\hat{\mathbf{B}} \right), \qquad \omega_B=\gamma B_0. \label{eq:Fz_zeeman} \ee 
Transforming to the magnet frame using $\mathbf{m}' = R\mathbf{m}$ gives 
\be F_{\mathrm Z} = -\omega_B R_{iz}^{(B)} R_{ij} m_j, \label{eq:Fz_zeeman_rot} \ee 
where Einstein summation over repeated indices is implied.
We then find \be F_{xx}^{(\mathrm Z)} = F_{yy}^{(\mathrm Z)} = F_{xy}^{(\mathrm Z)} = 0, \qquad F_z^{(\mathrm Z)} = -\omega_B R_{iz}^{(B)} R_{iz}. \label{eq:Fz_curvatures} \ee 

\textbf{Exchange energy.} In the long-wavelength limit, the exchange interaction is modeled as an isotropic contribution to the dispersion that scales as $k^2$. Evaluated at the equilibrium magnetization direction, we obtain \be F_{xx}^{(\mathrm{ex})} = F_{yy}^{(\mathrm{ex})} = Dk^2, \qquad F_{xy}^{(\mathrm{ex})} = 0, \qquad F_z^{(\mathrm{ex})} = 0, \label{eq:Fex} \ee where \be D = \frac{2\gamma A_{\mathrm{ex}}}{M_s} \ee is the exchange stiffness.

\textbf{Dipolar energy.} The dipolar tensor in $k$-space is given by~\cite{bertelli_magnetic_2020} \be \Gamma'(\mathbf{k}) = \begin{pmatrix} c^2 f_t & cs f_t & 0 \\ cs f_t & s^2 f_t & 0 \\ 0 & 0 & 1-f_t \end{pmatrix}, \label{eq:dipolar_tensor} \ee with $ c=\cos\phi_k$ , $s=\sin\phi_k $ and $ f_t(k) = 1-\frac{1-e^{-kt}}{kt}, \label{eq:ft} $ where $\phi_k$ is the azimuthal angle of the in-plane wave vector $\mathbf{k}$. The dipolar contribution to the effective field in the laboratory frame is \be H_i^{(\mathrm{dip})} = -\omega_M \Gamma_{ij}'(\mathbf{k}) m_j', \qquad \omega_M = \gamma\mu_0 M_s, \label{eq:Hdip} \ee with $M_s$ the saturation magnetization and $\mu_0$ the vacuum permeability. 

Transforming to the magnet frame via \be \Gamma(\mathbf{k}) = R^T \Gamma'(\mathbf{k}) R, \label{eq:Gamma_transform} \ee yields \be \begin{aligned} F_{xy}^{(\mathrm{dip})} &= -\omega_M \Gamma_{12}(\mathbf{k}), \\ F_{xx}-F_z &= \omega_M \left[ \Gamma_{33}(0)-\Gamma_{11}(\mathbf{k}) \right], \\ F_{yy}-F_z &= \omega_M \left[ \Gamma_{33}(0)-\Gamma_{22}(\mathbf{k}) \right]. \end{aligned} \label{eq:Fdip} \ee

\textbf{Uniaxial anisotropy.} The uniaxial anisotropy energy is modeled by \be F_{\mathrm u} = \gamma B_u \left(m_z'\right)^2. \label{eq:Fu} \ee Note that a negative (positive) value of $B_u$ corresponds to an out-of-plane easy (hard) axis. Transforming to the magnet frame using \be m_z' = R_{zi}m_i \ee gives \be F_{\mathrm u} = \gamma B_u R_{zi}R_{zj}m_im_j. \label{eq:Fu_rot} \ee Expanding around the equilibrium configuration $(m_x,m_y,m_z)=(0,0,1)$ yields \be \begin{aligned} F_{xx}^{(\mathrm u)} &= 2\gamma B_u R_{zx}^2, & F_{yy}^{(\mathrm u)} &= 2\gamma B_u R_{zy}^2, \\ F_{xy}^{(\mathrm u)} &= 2\gamma B_u R_{zx}R_{zy}, & F_{z}^{(\mathrm u)} &= 2\gamma B_u R_{zz}^2. \end{aligned} \label{eq:Fu_components} \ee

\textbf{Cubic anisotropy.} For a cubic crystal, the anisotropy energy in the crystal frame is modeled by \be F_{\mathrm{cub}} = \frac{1}{2}\gamma B_c \sum_{i=1}^{3} n_i^{4}, \label{eq:Fcub} \ee where $n_i$ are the magnetization components in the crystal frame. We rotate to the magnet frame via \be n_i = Q_{ij}m_j, \qquad Q = R^{(c)}R, \label{eq:Qmatrix} \ee where $R^{(c)}$ maps the laboratory frame to the crystal frame and $R$ maps the magnet frame to the laboratory frame. Expanding the energy to second order in $(m_x,m_y)$ around the equilibrium configuration $(0,0,1)$ yields \be \begin{aligned} F_{xx}^{(\mathrm{cub})} &= 6\gamma B_c \sum_i Q_{iz}^{2}Q_{ix}^{2}, & F_{yy}^{(\mathrm{cub})} &= 6\gamma B_c \sum_i Q_{iz}^{2}Q_{iy}^{2}, \\ F_{xy}^{(\mathrm{cub})} &= 6\gamma B_c \sum_i Q_{iz}^{2}Q_{ix}Q_{iy}, & F_{z}^{(\mathrm{cub})} &= 2\gamma B_c \sum_i Q_{iz}^{4}. \end{aligned} \label{eq:Fcub_components} \ee 

Substituting all contributions $F_{ij}^{(l)}$ into Eq.~\ref{eq:smit_beljers} and using the parameters for YIG listed in Table~S1 yields the spin-wave dispersion.

\textbf{Superconducting screening currents.} 
We now include the effect of the superconductor in the spin-wave dispersion. The stray field of the spin waves induces eddy currents in the metal, which in turn generate a magnetic field $\mathbf{B}_{\mathrm{SC}}'$ acting back on the magnetization. 

In the laboratory frame, we have~\cite{borst_observation_2023} \be \gamma \mathbf{B}_{\mathrm{SC}}' = i\omega \alpha_m(k,\omega) M_e'(\phi_k) \mathbf{m}', \label{eq:Bsc_lab} \ee where $\alpha_m$ is the dimensionless coupling coefficient characterizing the superconducting response. In the kinetic-inductance-dominated regime it reduces to \be \alpha_m = i\frac{\omega_{\mathrm{L}}}{\omega}. \label{eq:alpha_m} \ee Here, $\omega_{\mathrm{L}}$ is given by \be \omega_{\mathrm{L}} = \frac{\gamma \mu_0 M_s t}{2k\lambda_{\mathrm{L}}^2} g_t^2 \frac{1-e^{-2\kappa h}} {a_+^2-a_-^2e^{-2\kappa h}}. \label{eq:omegaL} \ee In this expression, $\lambda_{\mathrm{L}}$ is the London penetration depth, $ g_t = \frac{1-e^{-kt}}{kt}, \label{eq:gt} $ accounts for averaging over the film thickness, $h$ is the thickness of the superconducting layer, and $ a_\pm = 1 \pm \frac{\kappa}{k}. \label{eq:apm} $ The screening parameter $\kappa$ is \be \kappa = k \sqrt{ 1+\frac{1}{k^2\lambda_{\mathrm{L}}^2(T)} }. \label{eq:kappa} \ee Lastly, $M_e'(\phi_k)$ is a geometry-dependent coupling matrix given by 
\be M_e'(\phi_k) = \begin{pmatrix} \cos^2\phi_k & \sin\phi_k\cos\phi_k & i\cos\phi_k \\ \sin\phi_k\cos\phi_k & \sin^2\phi_k & i\sin\phi_k \\ -i\cos\phi_k & -i\sin\phi_k & 1 \end{pmatrix}. \label{eq:Me} \ee
Expressing this in the magnet frame, we obtain \be \gamma \mathbf{B}_{\mathrm{SC}} = i\omega \alpha_m(k,\omega) R^T M_e' R\,\mathbf{m} = D(\omega,k)\mathbf{m}, \label{eq:Bsc_magnetframe} \ee where we define the tensor $D_{ij}$. Incorporating this contribution into Eq.~\ref{eq:LL}, we obtain \be \begin{pmatrix} \dot{m}_x \\ \dot{m}_y \end{pmatrix} = \gamma \begin{pmatrix} -(F_{yx}-D_{yx}) & -(F_{yy}-D_{yy}-F_z) \\ (F_{xx}-D_{xx}-F_z) & (F_{xy}-D_{xy}) \end{pmatrix} \begin{pmatrix} m_x \\ m_y \end{pmatrix}. \label{eq:LL_SC} \ee Introducing the effective curvature coefficients \be F_{ij}^{\mathrm{eff}} = F_{ij} - D_{ij}, \label{eq:Feff} \ee the spin-wave dispersion becomes \be \omega = \frac{F_{xy}-F_{yx}}{2} + \sqrt{ \left(F_{xx}^{\mathrm{eff}}-F_z\right) \left(F_{yy}^{\mathrm{eff}}-F_z\right) - F_{xy}^{\mathrm{eff}} F_{yx}^{\mathrm{eff}} + \frac{\left(F_{xy}-F_{yx}\right)^2}{4} }, \label{eq:dispersion_SC} \ee and \be f_{\mathrm{sw}} = \frac{\omega}{2\pi}. \label{eq:fsw} \ee

\subsection{Fitting of the uniaxial anisotropy and London penetration depth}
In this section, we describe how we determine $B_u^{\mathrm{YIG}}$, $B_u^{\mathrm{YIG/SC}}$, and $\lambda_{\mathrm{L}}$ from the spin-wave vectors extracted from the spatial spin-wave maps, as described above, using the dispersion relation derived in Eq.~\ref{eq:dispersion_SC}. For the $45^\circ$ device shown in Fig.~2 of the main text, we determine $B_u^{\mathrm{YIG}}$ by minimizing 
\be \chi^2 = \sum_{\mathbf{k}_{\mathrm{YIG}}} \left[ f_{\mathrm{sw}} \bigl( \mathbf{k}_{\mathrm{YIG}}, B_u^{\mathrm{YIG}}, \mathbf{B}, p \bigr) - f_{\mathrm{ESR}} \right]^2, \label{eq:chi2_YIG} \ee 
where the sum runs over all extracted wave vectors at the different temperatures, $B_u^{\mathrm{YIG}}$ is the uniaxial anisotropy field (the only free parameter), $p$ denotes the independently determined parameters listed in Table~S1, and $\mathbf{B}$ is the applied magnetic field. From this procedure, we obtain 
$ B_u^{\mathrm{YIG}} = -6.7 \pm 1~\mathrm{mT}$. 
This value is approximately three times larger than the room-temperature value reported for thin-film YIG~\cite{dubs_low_2020}.

\begin{figure*}[!ht]
\includegraphics{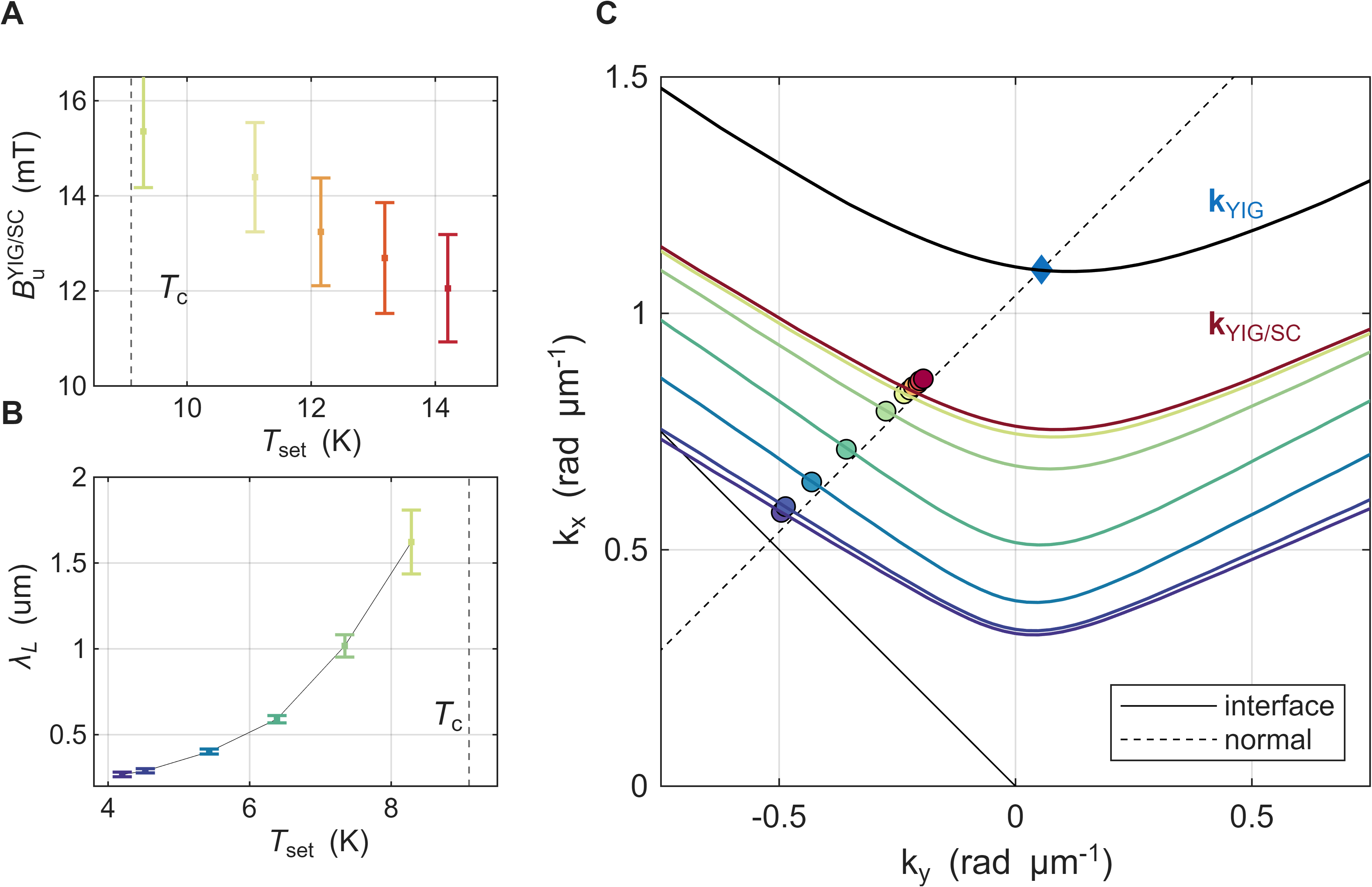}
\caption{ \textbf{Fitting of uniaxial anisotropy and London penetration depth as a function of temperature.} (\textbf{A}) Extracted uniaxial anisotropy field $B_u^{\mathrm{YIG/SC}}$ as a function of the set temperature for the device with a $45^\circ$ interface, using data acquired above the superconducting transition temperature $T_c$. (\textbf{B}) Extracted London penetration depth $\lambda_{\mathrm{L}}$ as a function of set temperature using $B_u^{\mathrm{YIG/SC}} = 15~\mathrm{mT}$, as extracted just above $T_c$ in (\textbf{A}). (\textbf{C}) Isofrequency contours of the spin-wave dispersion together with the extracted wave vectors. The coordinate system is rotated by $45^\circ$ with respect to the real-space maps such that $k_y$ is aligned with the microstrip direction. Circles and diamonds indicate the measured wave vectors $\mathbf{k}_{\mathrm{YIG/SC}}$ and $\mathbf{k}_{\mathrm{YIG}}$, respectively. Solid lines show the corresponding isofrequency contours computed using the fitted parameters. The color scale indicates $T_{\mathrm{set}}$ as shown in (\textbf{A}) and (\textbf{B}). The wave vectors and experimental conditions are the same as those in Fig.~\ref{fig:S1} and Fig.~2 of the main text. } \label{fig:S2}
\end{figure*}

The wave vector $\mathbf{k}_{\mathrm{YIG/SC}}$ exhibits a weak temperature dependence even above $T_c$, which we attribute to temperature-dependent strain in the YIG/NbTiN region affecting the uniaxial anisotropy. For each temperature above $T_c$, we determine $B_u^{\mathrm{YIG/SC}}$ directly by numerically solving \be f_{\mathrm{ESR}} = f_{\mathrm{sw}} \Bigl( \mathbf{k}_{\mathrm{YIG/SC}}(T), B_u^{\mathrm{YIG/SC}}(T) \Bigr). \label{eq:Bu_SC_fit} \ee This procedure yields the temperature dependence of $B_u^{\mathrm{YIG/SC}}(T)$ shown in Fig.~\ref{fig:S2}A.

To estimate the London penetration depth $\lambda_{\mathrm{L}}$, we fix $B_u^{\mathrm{YIG/SC}}$ to its value immediately above $T_c$ and attribute the change in wave vector below $T_c$ entirely to superconducting screening. Fixing $ B_u^{\mathrm{YIG/SC}} = 15 \pm 1~\mathrm{mT}$, 
as extracted from Fig.~\ref{fig:S2}A at $T_{\mathrm{set}} = 9.3~\mathrm{K}$, we determine $\lambda_{\mathrm{L}}$ by solving \be f_{\mathrm{ESR}} = f_{\mathrm{sw}} \Bigl( \mathbf{k}_{\mathrm{YIG/SC}}(T), B_u^{\mathrm{YIG/SC}}, \mathbf{B}, p, \lambda_{\mathrm{L}} \Bigr). \label{eq:lambda_fit} \ee The resulting temperature dependence of the London penetration depth is shown in Fig.~\ref{fig:S2}B. The corresponding isofrequency contours for the extracted $B_u^{\mathrm{YIG}}$, $B_u^{\mathrm{YIG/SC}}$, and $\lambda_{\mathrm{L}}$ are compared with the measured $\mathbf{k}_{\mathrm{YIG}}$ and $\mathbf{k}_{\mathrm{YIG/SC}}$ in Fig.~\ref{fig:S2}C.

The uncertainties in the fitted parameters are estimated using a Monte Carlo approach, in which $\mathbf{k}_{\mathrm{YIG/SC}}$ and the applied magnetic field are varied within their experimental uncertainties assuming Gaussian distributions. For each realization, the fitting procedure is repeated, and the reported uncertainty is taken as the standard deviation of the resulting parameter distribution. Each distribution is typically obtained from 200 realizations. 

\begin{figure*}[!ht]
\includegraphics{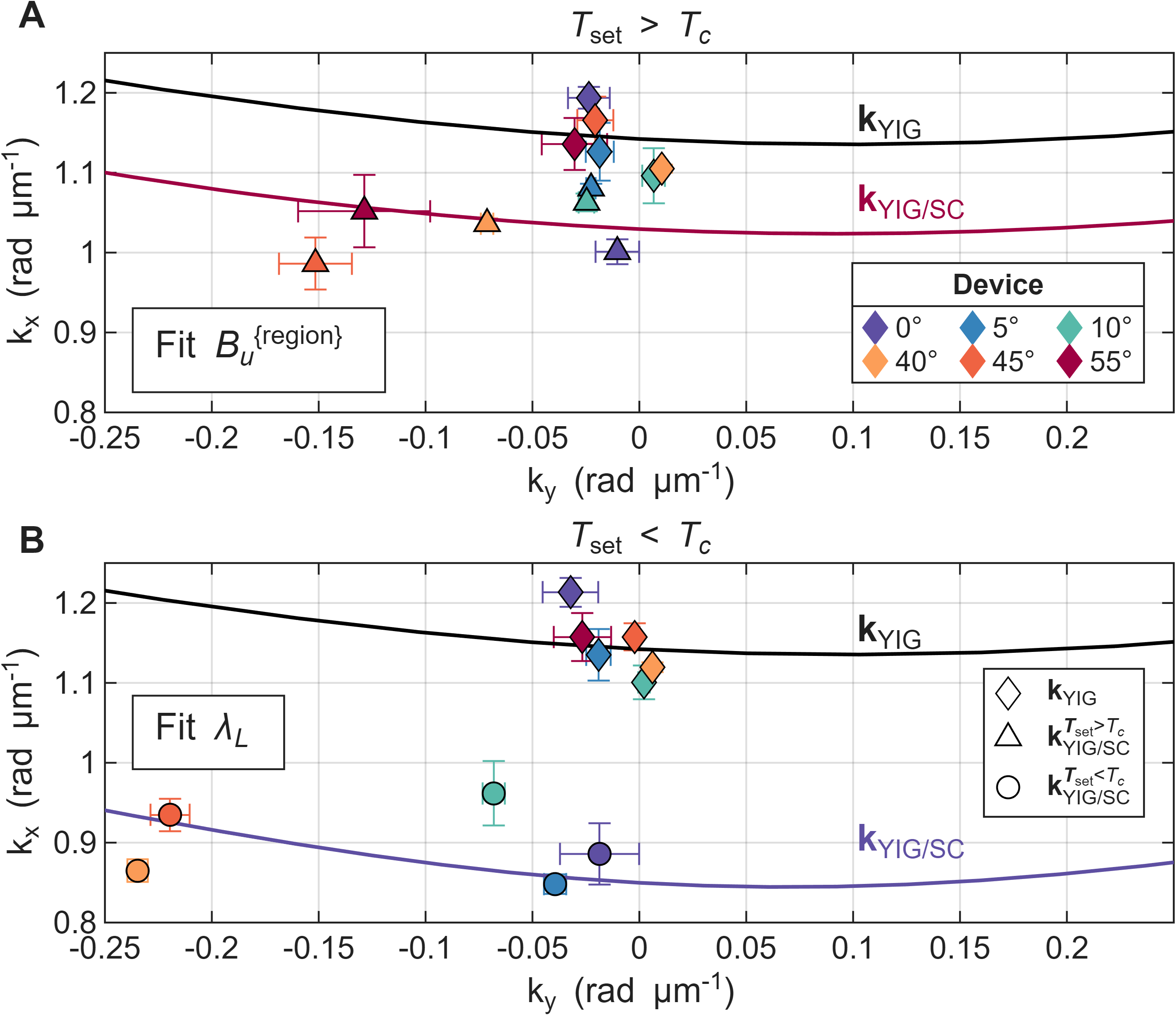}
\caption{ \textbf{Fitting uniaxial anisotropy and London penetration depth for different interface angles.} (\textbf{A}) Extracted wave vectors for the YIG and YIG/SC regions at temperatures above the superconducting transition temperature. Solid lines show the calculated isofrequency contours obtained using the fitted uniaxial anisotropy parameters $B_u^{\mathrm{region}}$. (\textbf{B}) Extracted wave vectors for the YIG and YIG/SC regions at temperatures below $T_c$ ($T_{\mathrm{set}} = 4$--$5~\mathrm{K}$). Solid lines show the isofrequency contours computed using the fitted London penetration depth $\lambda_{\mathrm{L}}$. Measurement parameters are $B_{\mathrm{ip}} = 8.0~\mathrm{mT}$, $f_{\mathrm{ESR}} = 2.603~\mathrm{GHz}$, and $\lambda_{\mathrm{L}} = 800~\mathrm{nm}$. } \label{fig:S3}
\end{figure*}

For the data shown in Fig.~3 of the main text, a similar fitting procedure is applied. Because these measurements correspond to different superconducting elements, $B_u^{\mathrm{YIG}}$ and $B_u^{\mathrm{YIG/SC}}$ show a distribution across devices. We therefore perform a least-squares minimization over all datasets acquired above $T_c$, yielding $ B_u^{\mathrm{YIG}} = 1.5 \pm 2~\mathrm{mT}$ and $  B_u^{\mathrm{YIG/SC}} = 10.5 \pm 4~\mathrm{mT}$.  The sign change in $B_u$ has previously been observed in FMR absorption measurements at low temperatures~\cite{serha_magnetic_2024}. Using these parameters, we determine the London penetration depth for the dataset and obtain $\lambda_{\mathrm{L}} = 800 \pm 150~\mathrm{nm}. $ These values are used to calculate the theoretical curve shown in Fig.~3A-B of the main text. In Fig.~\ref{fig:S3}, we compare the extracted $\mathbf{k}_{\mathrm{YIG}}$ and $\mathbf{k}_{\mathrm{YIG/SC}}$ with the corresponding calculated isofrequency contours.

\subsection{Electrical measurement of the superconducting transition temperature} 
\begin{figure*}[!ht]
\includegraphics{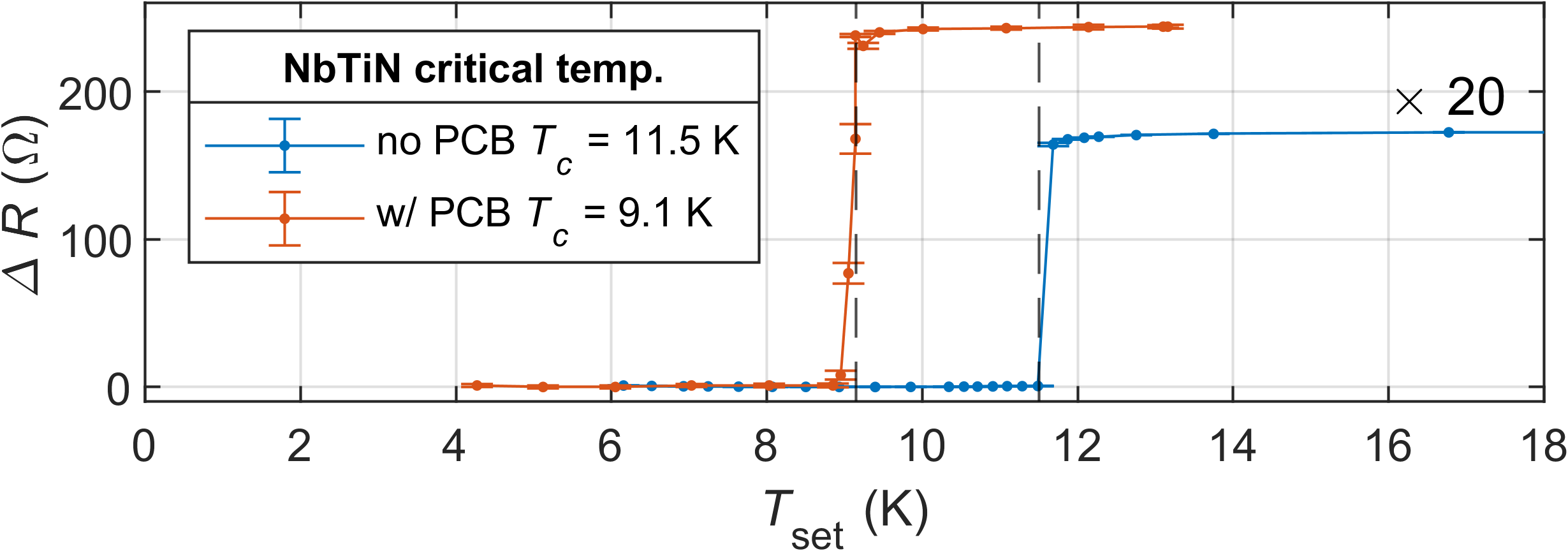}
\caption{ \textbf{Two-probe electrical measurements of the superconducting transition.} Electrical resistance as a function of set temperature for two different sample-loading configurations: a sample mounted on a PCB (the sample used in the main manuscript) and a reference sample mounted without a PCB. For the reference sample (no PCB), a current of $10~\mathrm{mA}$ is applied to a Si chip coated with a $\sim4~\upmu\mathrm{m}$-thick NbTiN film. For the PCB-mounted sample, a current of $10~\upmu\mathrm{A}$ is applied across a bowtie-shaped structure ($150~\mathrm{nm}$ thick with a minimum width of $5~\upmu\mathrm{m}$). No dependence on the applied RF power ($<12~\mathrm{dBm}$) is observed. } \label{fig:S4}
\end{figure*}

The critical temperature of the NbTiN was determined using two-probe transport measurements (Fig.~\ref{fig:S4}). We found that the measured critical temperature depended on the thermal coupling between the sample and its mount. For all measurements presented in the main text, the sample was mounted onto a printed circuit board (PCB), which was in turn mounted on a copper cold finger. This intermediate PCB layer led to a reduced apparent $T_c$ compared to that of a separate NbTiN chip that was attached directly to the cold finger (Fig.~\ref{fig:S4}). As such, we expect the sample temperature in our measurements to deviate by approximately $2~\mathrm{K}$ from the cryostat set temperature. This offset does not affect our analysis. 

\subsection{Effect of global heating} 
\begin{figure*}[!ht]
\includegraphics{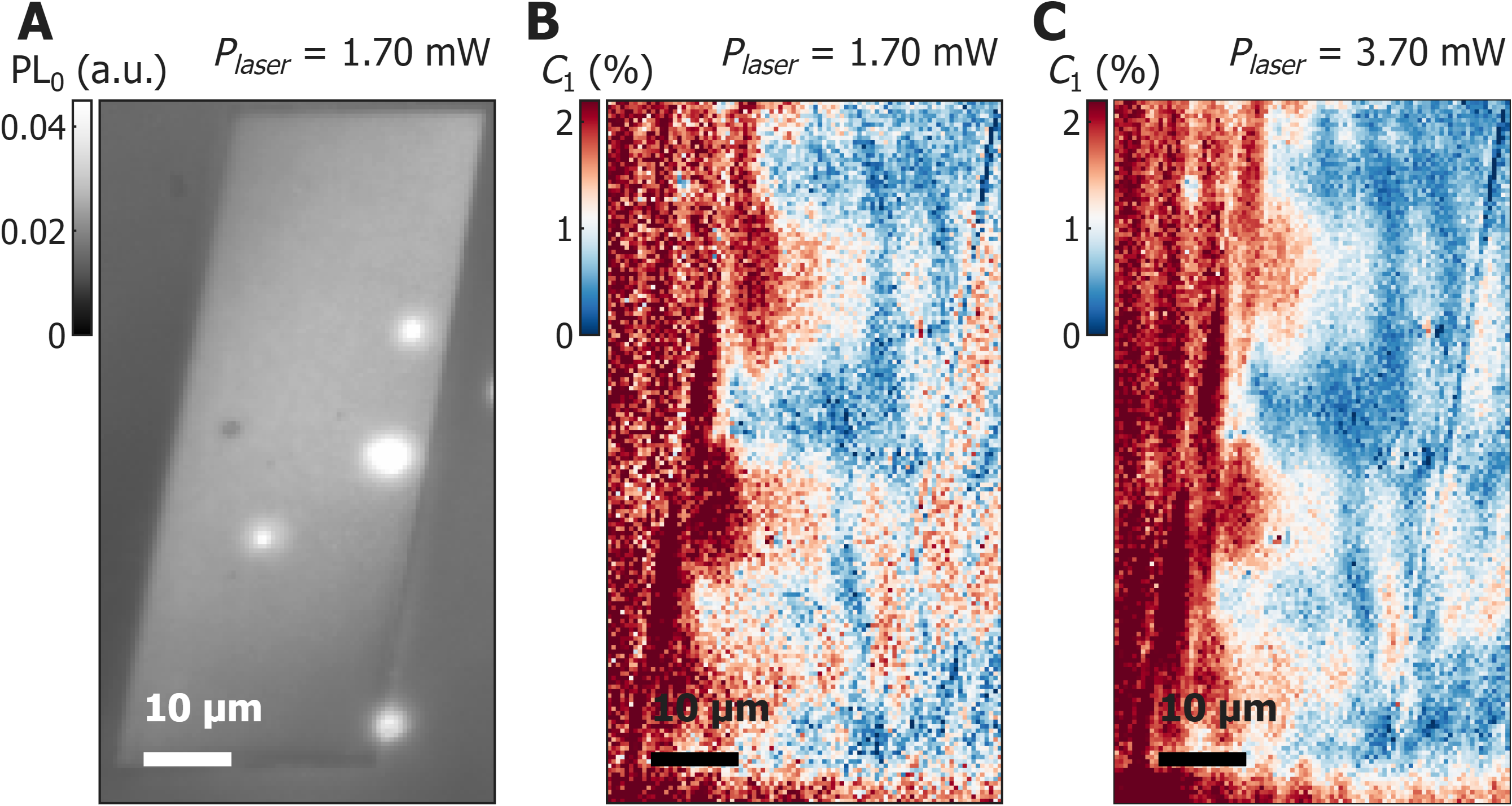}
\caption{ \textbf{Spin-wave maps as a function of laser power.} (\textbf{A}) Spatial map of the NV photoluminescence in the absence of microwave excitation, defining the device geometry. The interface is oriented at an angle of $10^\circ$ with respect to the microstrip. Bright spots correspond to pits in the diamond membrane formed during the etching process, where the NV photoluminescence is locally enhanced. (\textbf{B, C}) Spatial maps of the NV electron spin resonance contrast, $C_1$, for two different laser powers at base temperature ($T_{\mathrm{set}} = 4.6~\mathrm{K}$). Spin waves are excited at a frequency of $f_{\mathrm{ESR}} = 2.663~\mathrm{GHz}$ and an in-plane magnetic field of $B_{\mathrm{ip}} = 6.5~\mathrm{mT}$. } \label{fig:S5}
\end{figure*}

\textbf{Laser power.} To investigate possible global heating effects induced by the laser, we acquired spin-wave maps as a function of laser power (Fig.~\ref{fig:S5}). From these maps, we observe that laser powers above approximately $6~\mathrm{mW}$ lead to noticeable changes in the refraction patterns compared to lower powers, consistent with heating of the superconducting layer. In the measurements presented in the main manuscript, the laser power was therefore kept at $3~\mathrm{mW}$, corresponding to approximately $100~\upmu\mathrm{W}$ at the sample due to overfilling of the objective. 

\begin{figure*}[!ht]
\includegraphics{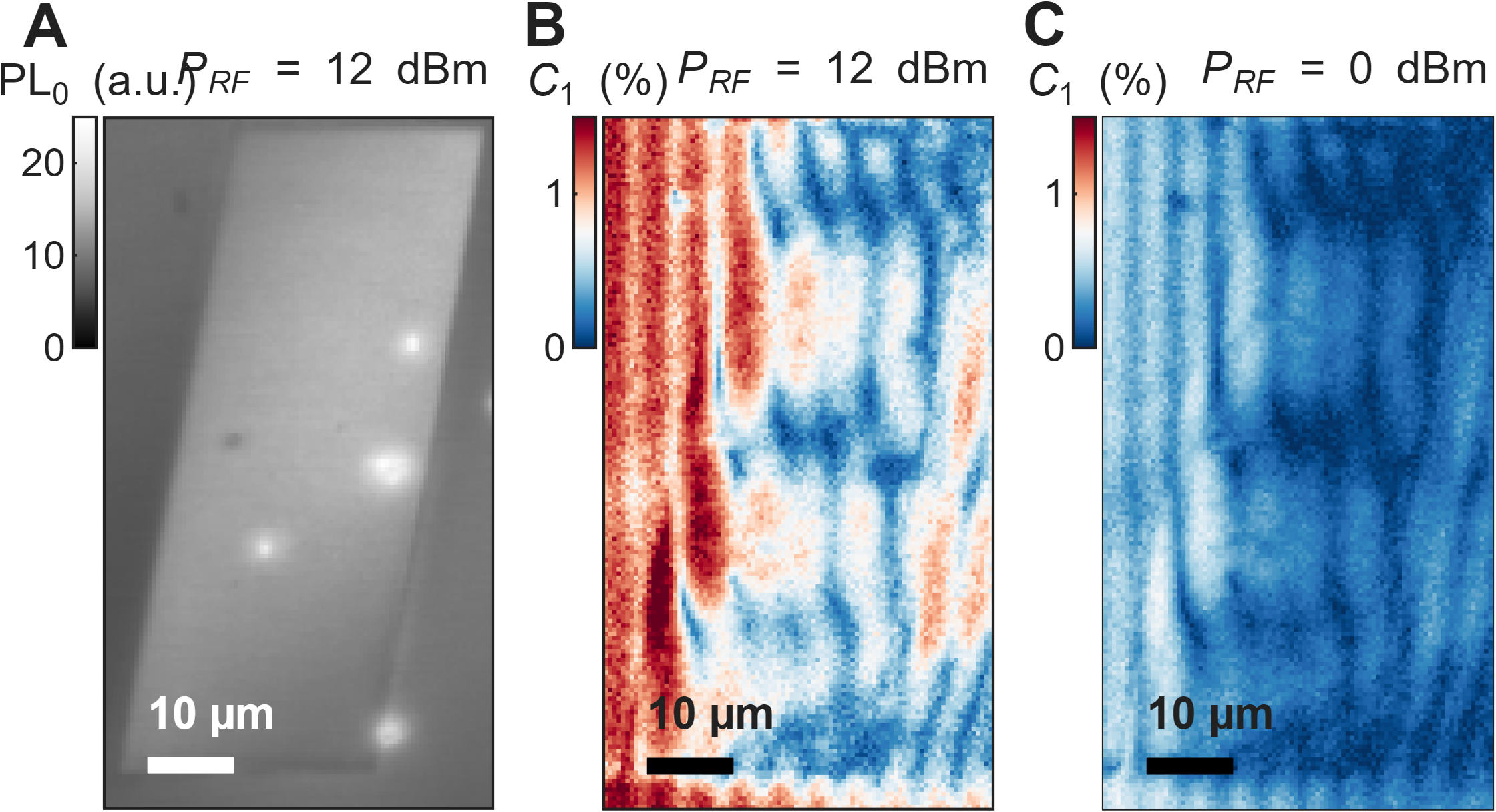}
\caption{ \textbf{Spin-wave maps as a function of microwave power.} (\textbf{A}) Spatial map of the NV photoluminescence in the absence of microwave excitation, defining the device geometry. The interface is oriented at an angle of $10^\circ$ with respect to the microstrip. (\textbf{B, C}) Spatial maps of the NV electron spin resonance contrast, $C_1$, for two different RF powers at base temperature ($T_{\mathrm{set}} = 4.6~\mathrm{K}$). Spin waves are excited at a frequency of $f_{\mathrm{ESR}} = 2.663~\mathrm{GHz}$ and an in-plane magnetic field of $B_{\mathrm{ip}} = 6.5~\mathrm{mT}$. } \label{fig:S6}
\end{figure*}

\textbf{Microwave power.} A similar control experiment was performed by varying the applied microwave (MW) power (Fig.~\ref{fig:S6}). We find that reducing the MW power by $12~\mathrm{dB}$ does not lead to observable changes in the spin-wave maps, indicating that the MW excitation does not significantly affect the measured spin-wave propagation within the explored power range.

\subsection{Additional Results} \textbf{Manuscript Fig.~2.} In the main paper, we plot $\theta_{\mathrm{YIG/SC}}$ as a function of temperature. The corresponding spin-wave maps are shown in Fig.~\ref{fig:S1}, while the fitted London penetration depths and uniaxial anisotropy parameters are shown in Fig.~\ref{fig:S2}.

\textbf{Manuscript Fig.~3.} The datasets shown in Fig.~\ref{fig:S10}--\ref{fig:S15} were used to extract the refracted wave-vector angles above and below the superconducting transition temperature. 

\textbf{Manuscript Fig.~4.} The spin-wave vectors $\mathbf{k}_{\mathrm{YIG}}$ and $\mathbf{k}_{\mathrm{YIG/SC}}$ shown in Fig.~4E of the main text are extracted from the two-dimensional Fourier transforms of the spatial maps shown in Fig.~\ref{fig:S8} and Fig.~\ref{fig:S9}. In these datasets, we observe additional peaks in both the Fourier spectra and the corresponding NV ESR contrast maps. The origin of these additional modes is not fully understood, but similar features are consistently observed across multiple datasets.

\begin{figure*}[!ht]
\includegraphics[]{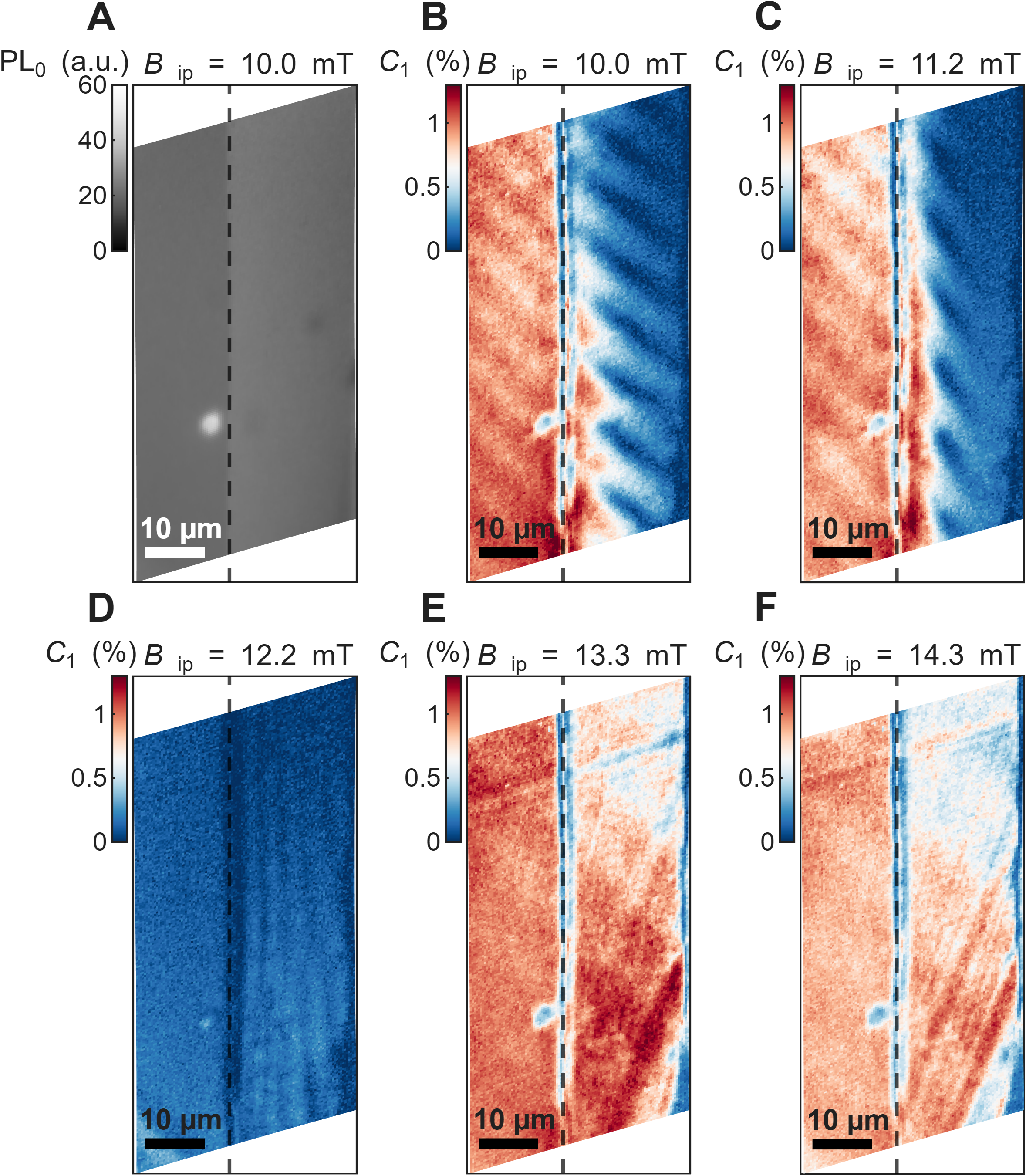}
\caption{ \textbf{Spin-wave maps at different applied magnetic fields above the superconducting transition.} (\textbf{A}) Spatial map of the NV photoluminescence in the absence of microwave excitation. The dashed line indicates the boundary between the YIG and YIG/NbTiN regions. The interface is oriented at $45^\circ$ with respect to the microstrip. For clarity, the maps are rotated, resulting in visible white regions. (\textbf{B--F}) Spatial maps of the NV electron spin resonance contrast, $C_1$, for five different magnetic-field values acquired at $T_{\mathrm{set}} = 11.1~\mathrm{K}$. } \label{fig:S7}
\end{figure*}

\begin{figure*}[!ht]
\includegraphics{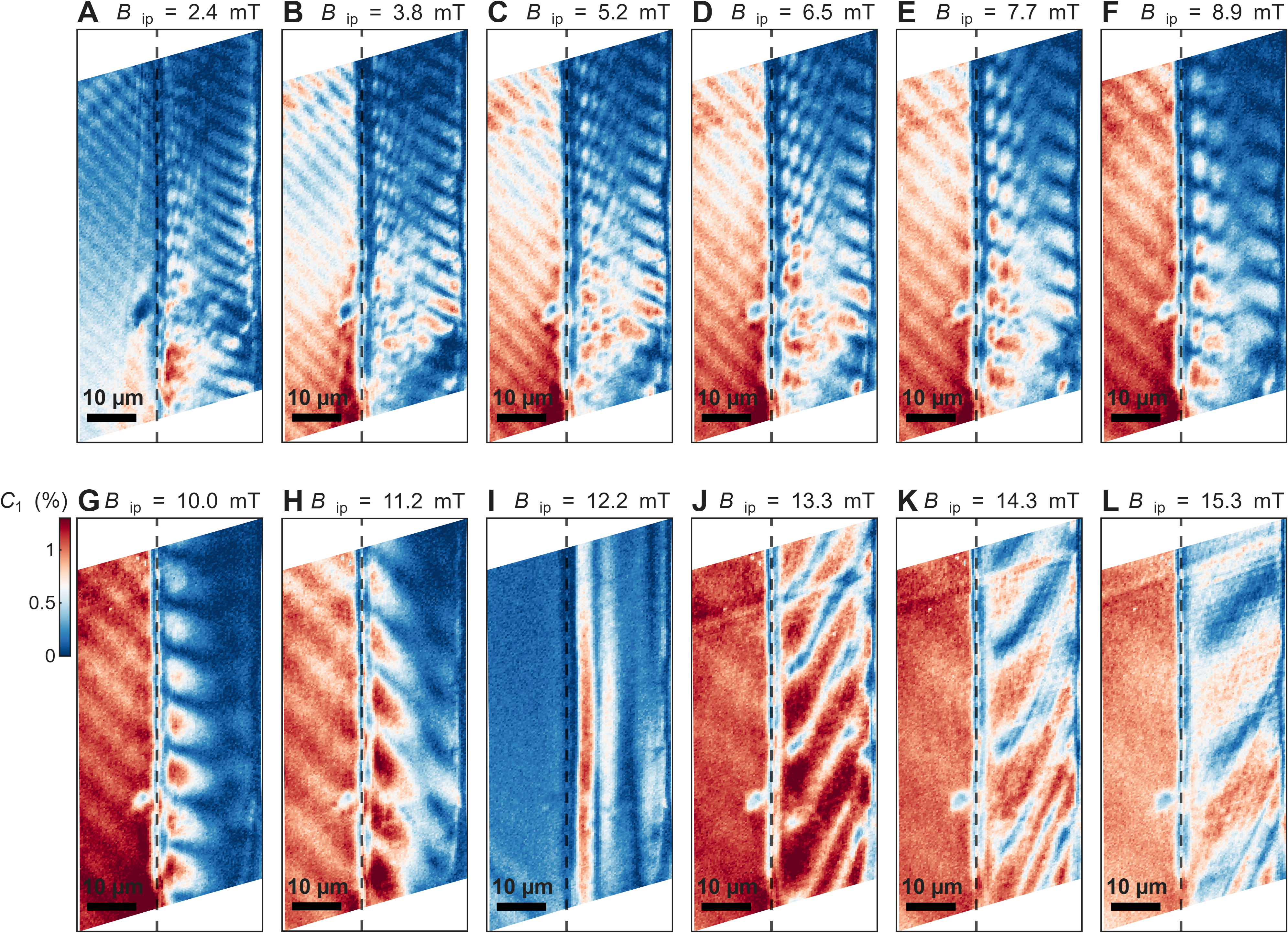}
\caption{ \textbf{Spin-wave maps at different applied magnetic fields below the superconducting transition.} (\textbf{A--L}) Spatial maps of the NV electron spin resonance contrast, $C_1$, acquired at different magnetic-field values and a set temperature of $T_{\mathrm{set}} = 4.4~\mathrm{K}$. These maps correspond to the measurements shown in Fig.~4A--C of the main manuscript. } \label{fig:S8}
\end{figure*}

\begin{figure*}[!ht]
\includegraphics{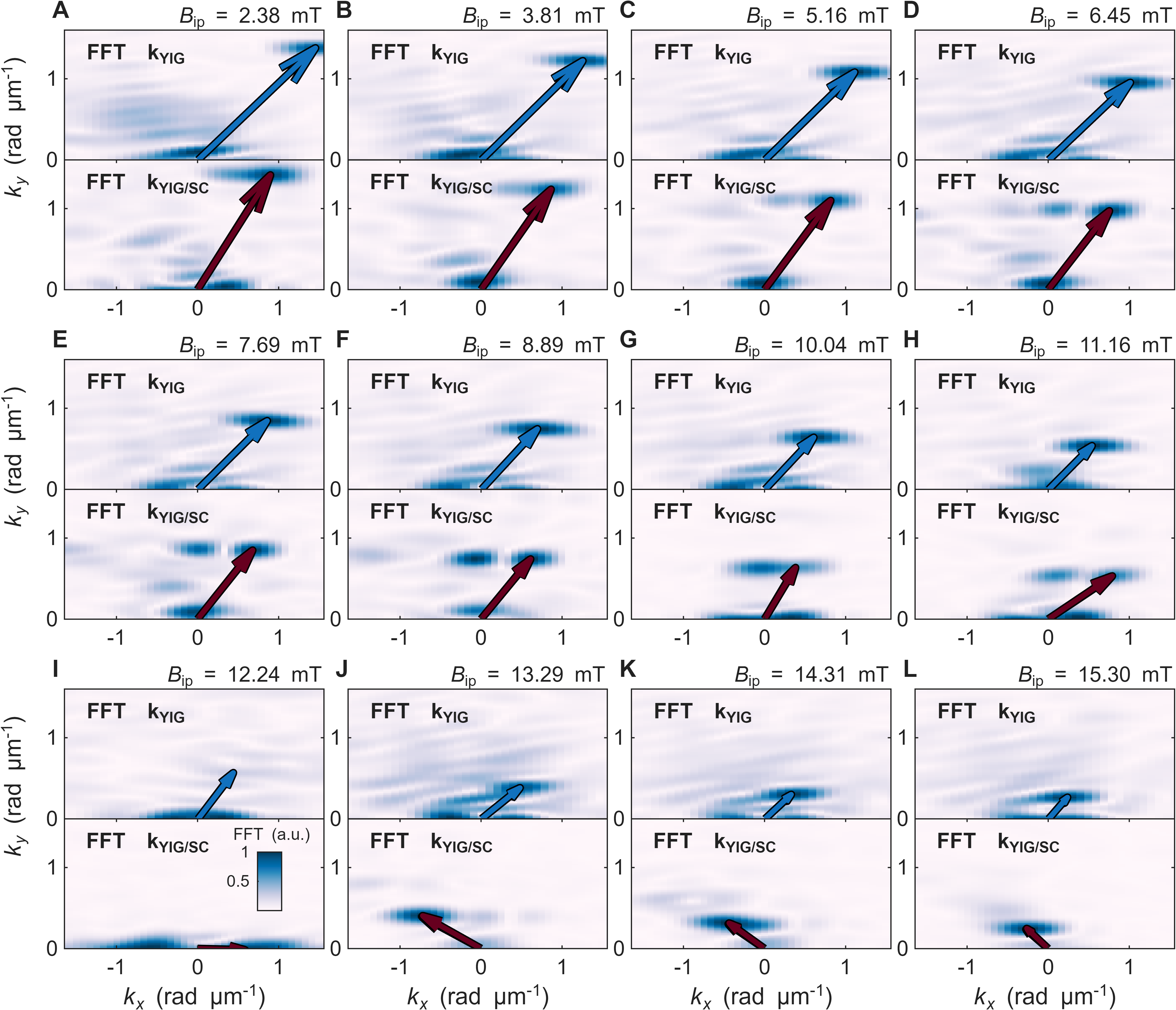}
\caption{ \textbf{Two-dimensional Fourier transforms of spin-wave maps acquired at different magnetic fields below the superconducting transition.} The top panels show the Fourier transforms of the YIG region, while the bottom panels show those of the YIG/NbTiN region. Arrows indicate the peak positions in Fourier space used to extract the spin-wave wave vectors shown in Fig.~4E of the main manuscript. Panel (\textbf{I}) corresponds to a case with negligible spin-wave excitation, for which the peak position cannot be reliably determined. All data were acquired at $T_{\mathrm{set}} = 4.4~\mathrm{K}$. } \label{fig:S9}
\end{figure*}

\begin{figure*}[!ht]
\includegraphics[width=\textwidth]{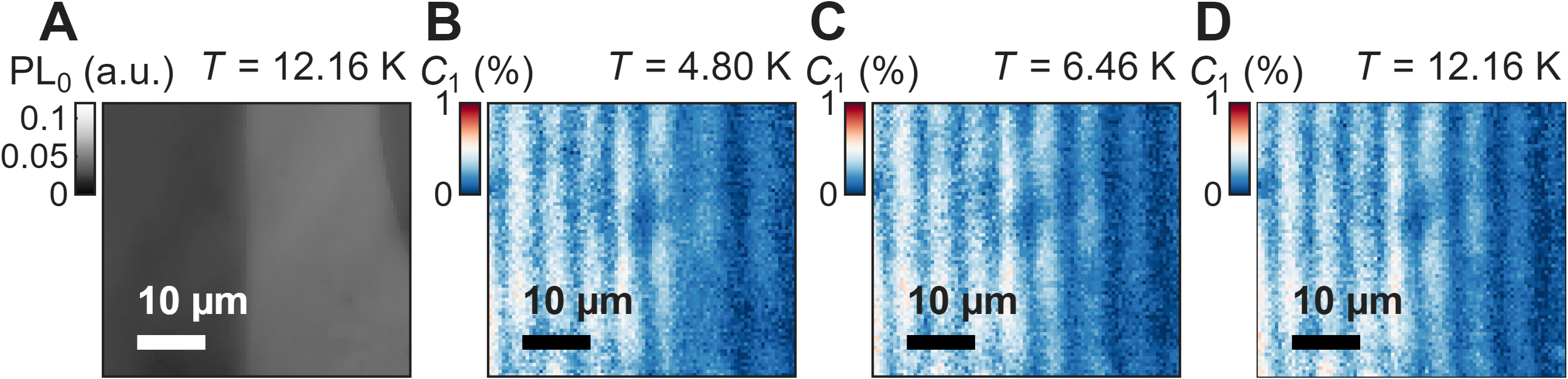}
\caption{ \textbf{Spin-wave maps for a straight ($0^\circ$) interface.} (\textbf{A}) Spatial map of the NV photoluminescence in the absence of microwave excitation. The light-grey region corresponds to the YIG/NbTiN structure. This panel shows a zoom-in of the lens element. (\textbf{B--D}) Spatial maps of the NV electron spin resonance contrast, $C_1$, acquired at different temperatures. Measurements are performed at an in-plane magnetic field of $B_{\mathrm{ip}} = 8.0~\mathrm{mT}$ and an excitation frequency of $f_{\mathrm{ESR}} = 2.603~\mathrm{GHz}$. } \label{fig:S10}
\end{figure*}

\begin{figure*}[!ht]
\includegraphics[width=\textwidth]{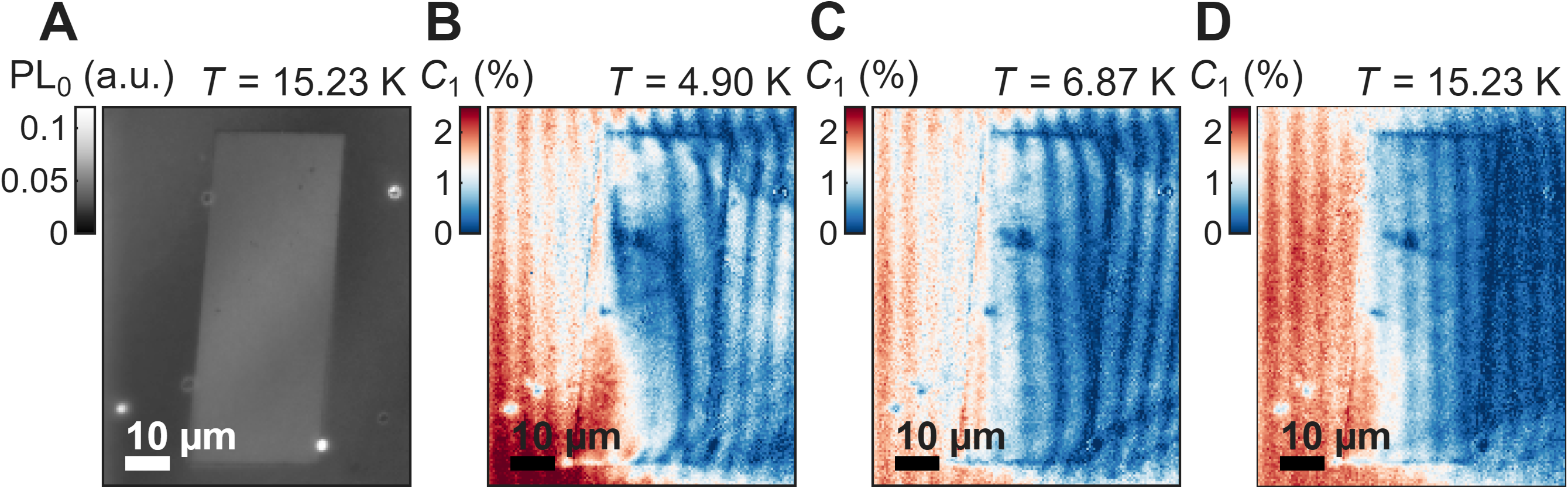}
\caption{ \textbf{Spin-wave maps for a $5^\circ$ interface.} (\textbf{A}) Spatial map of the NV photoluminescence in the absence of microwave excitation. The light-grey region corresponds to the YIG/NbTiN structure. (\textbf{B--D}) Spatial maps of the NV electron spin resonance contrast, $C_1$, acquired at different temperatures. Measurements are performed at an in-plane magnetic field of $B_{\mathrm{ip}} = 8.0~\mathrm{mT}$ and an excitation frequency of $f_{\mathrm{ESR}} = 2.603~\mathrm{GHz}$. } \label{fig:S11}
\end{figure*}

\begin{figure*}[!ht]
\includegraphics[width=\textwidth]{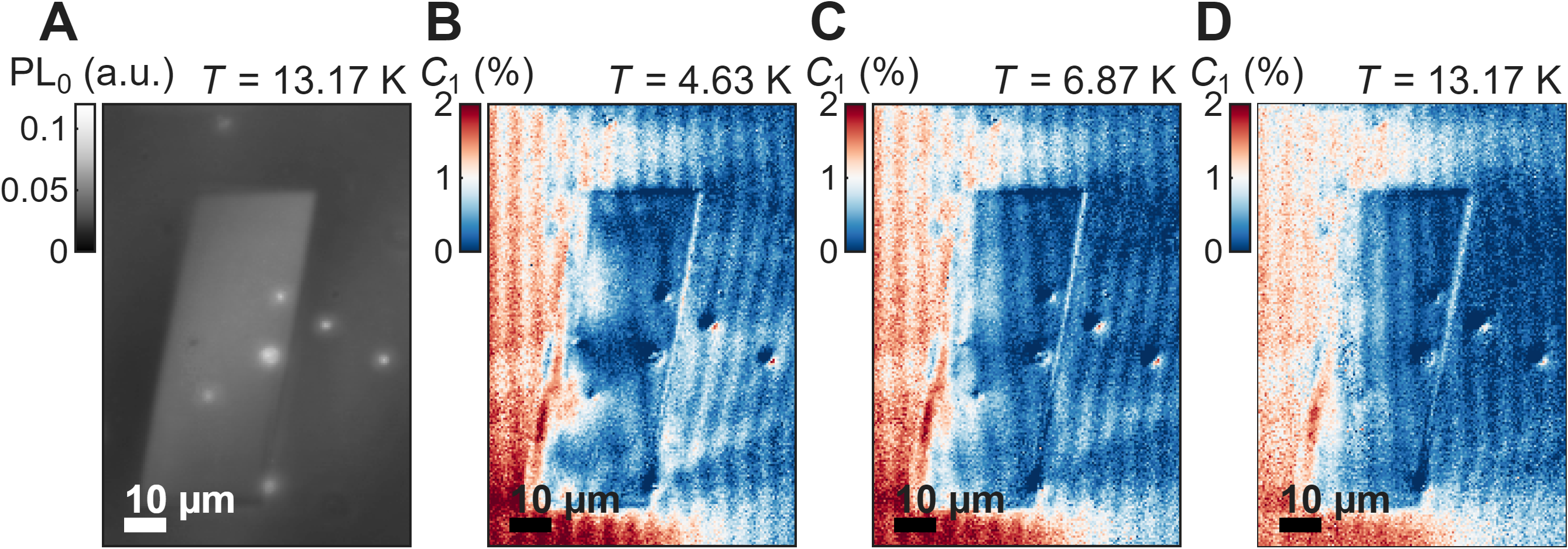}
\caption{ \textbf{Spin-wave maps for a $10^\circ$ interface.} (\textbf{A}) Spatial map of the NV photoluminescence in the absence of microwave excitation. The light-grey region corresponds to the YIG/NbTiN structure. (\textbf{B--D}) Spatial maps of the NV electron spin resonance contrast, $C_1$, acquired at different temperatures. Measurements are performed at an in-plane magnetic field of $B_{\mathrm{ip}} = 8.0~\mathrm{mT}$ and an excitation frequency of $f_{\mathrm{ESR}} = 2.603~\mathrm{GHz}$. } \label{fig:S12}
\end{figure*}

\begin{figure*}[!ht]
\includegraphics[width=\textwidth]{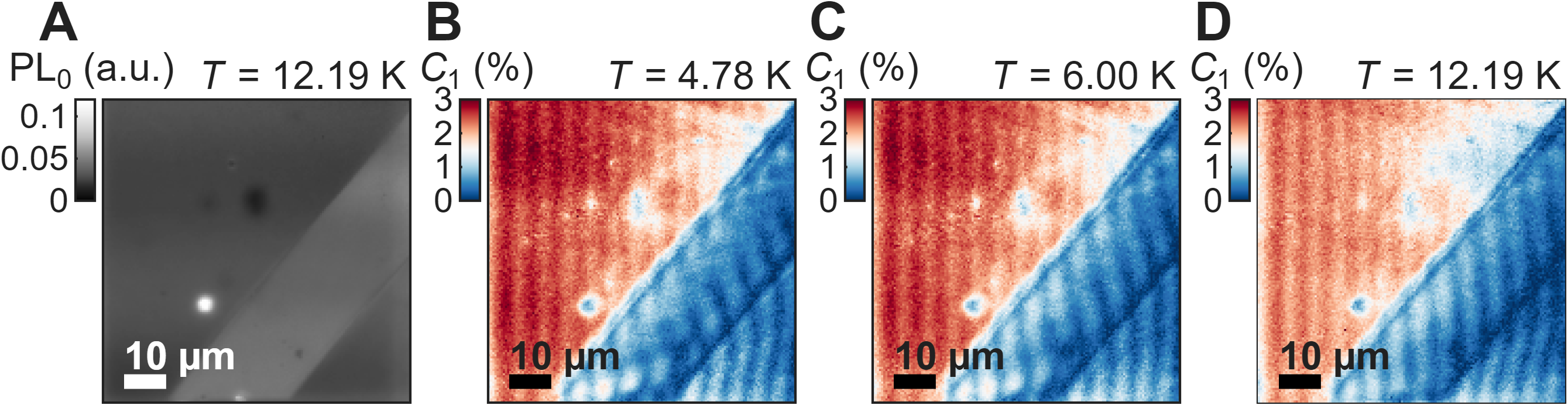}
\caption{ \textbf{Spin-wave maps for a $40^\circ$ interface.} (\textbf{A}) Spatial map of the NV photoluminescence in the absence of microwave excitation. The light-grey region corresponds to the YIG/NbTiN structure. (\textbf{B--D}) Spatial maps of the NV electron spin resonance contrast, $C_1$, acquired at different temperatures. Measurements are performed at an in-plane magnetic field of $B_{\mathrm{ip}} = 8.0~\mathrm{mT}$ and an excitation frequency of $f_{\mathrm{ESR}} = 2.603~\mathrm{GHz}$. } \label{fig:S13}
\end{figure*}

\begin{figure*}[!ht]
\includegraphics[width=\textwidth]{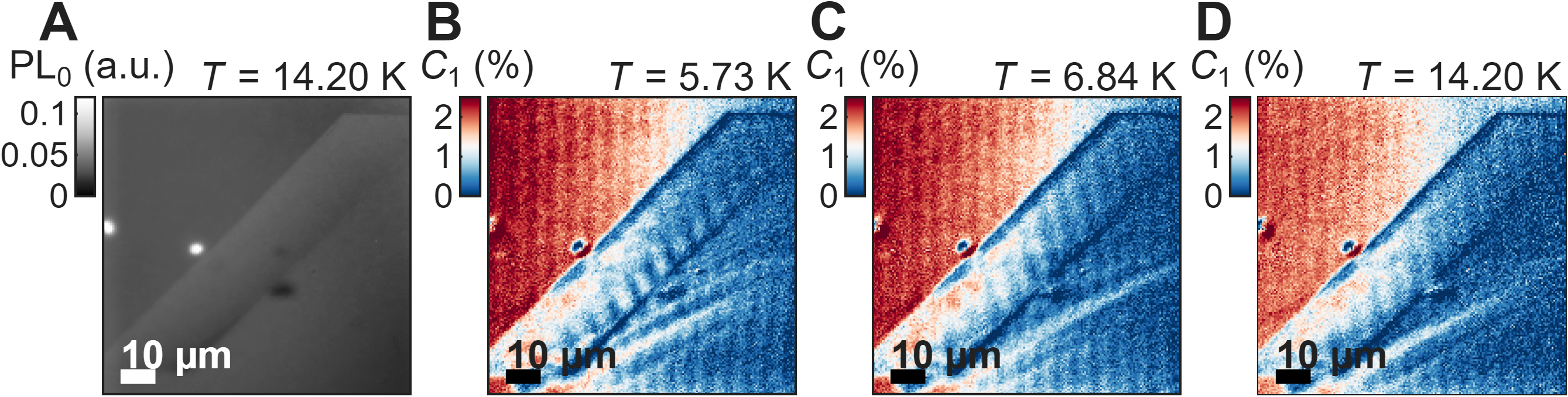}
\caption{ \textbf{Spin-wave maps for a $45^\circ$ interface.} (\textbf{A}) Spatial map of the NV photoluminescence in the absence of microwave excitation. The light-grey region corresponds to the YIG/NbTiN structure. (\textbf{B--D}) Spatial maps of the NV electron spin resonance contrast, $C_1$, acquired at different temperatures. Measurements are performed at an in-plane magnetic field of $B_{\mathrm{ip}} = 8.0~\mathrm{mT}$ and an excitation frequency of $f_{\mathrm{ESR}} = 2.603~\mathrm{GHz}$. } \label{fig:S14}
\end{figure*}

\begin{figure*}[!ht]
\includegraphics[width=\textwidth]{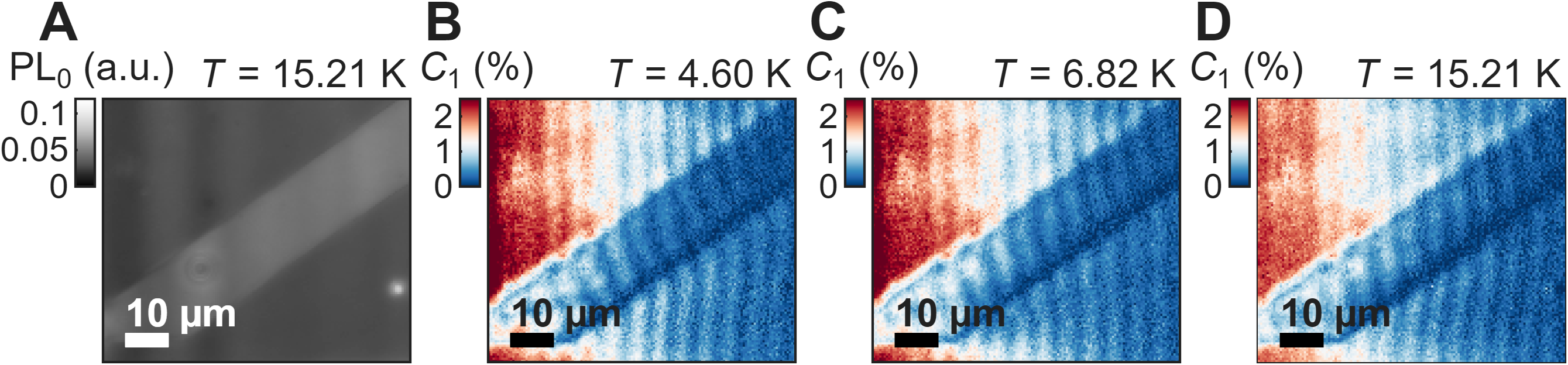}
\caption{ \textbf{Spin-wave maps for a $55^\circ$ interface.} (\textbf{A}) Spatial map of the NV photoluminescence in the absence of microwave excitation. The light-grey region corresponds to the YIG/NbTiN structure. (\textbf{B--D}) Spatial maps of the NV electron spin resonance contrast, $C_1$, acquired at different temperatures. Measurements are performed at an in-plane magnetic field of $B_{\mathrm{ip}} = 8.0~\mathrm{mT}$ and an excitation frequency of $f_{\mathrm{ESR}} = 2.603~\mathrm{GHz}$. } \label{fig:S15}
\end{figure*}

To further investigate the field dependence, we acquired spin-wave maps for a range of magnetic-field amplitudes above the superconducting transition temperature (Fig.~\ref{fig:S7}). In this regime, we find that for $B_{\mathrm{ip}}=12.2~\mathrm{mT}$ the spin-wave signal is strongly suppressed. We attribute this suppression to the overlap of the NV resonance frequency with the first-order perpendicular mode of the spin-wave dispersion, which reduces the excitation efficiency of the fundamental mode~\cite{kalinikos_theory_1986}.

In this field regime, direct inductive coupling between the microstrip and the NbTiN becomes visible, leading to spin-wave excitation in the YIG/NbTiN region with wave fronts oriented perpendicular to the interface~\cite{borst_observation_2023}. This behavior is not observed at other magnetic-field values. At higher fields, the reduced contribution of higher-order modes leads to lower damping and an increased NV contrast. For clarity, datasets acquired at $B_{\mathrm{ip}}=12.2~\mathrm{mT}$ are not included in the main manuscript. 
\subsection{Extra spin-wave maps.} 
\begin{figure*}[!ht]
\includegraphics[width=\textwidth]{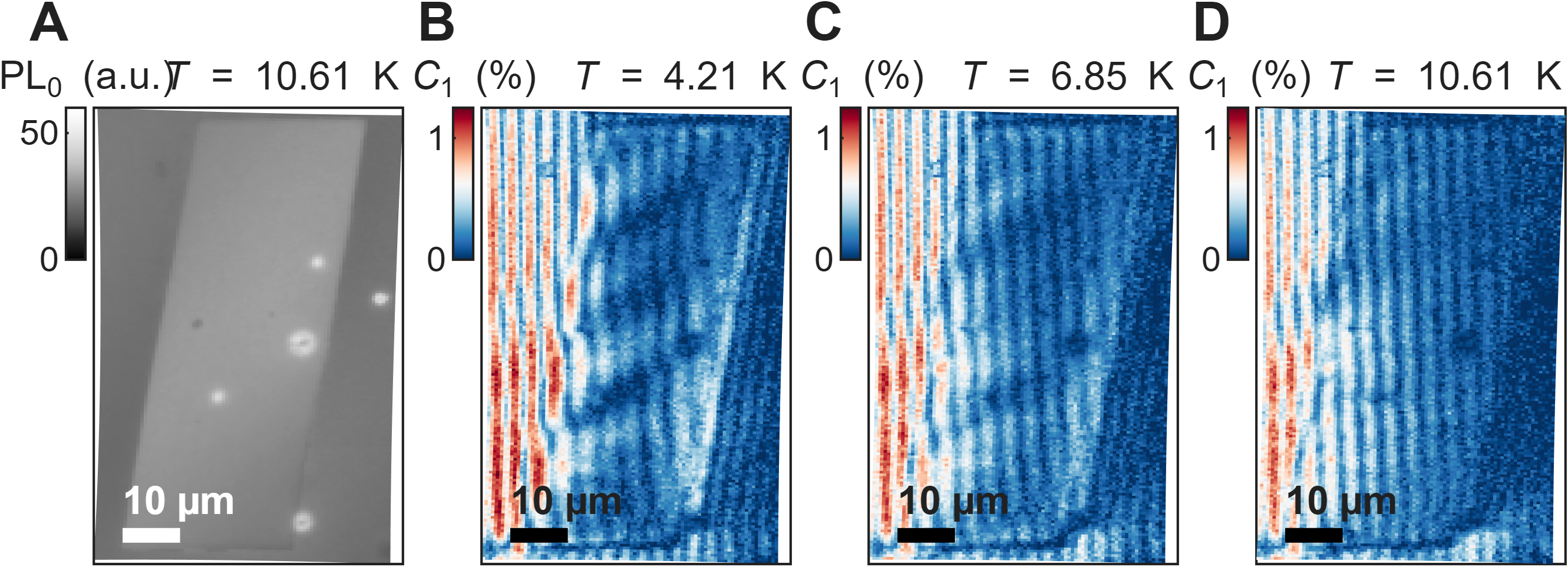}
\caption{ \textbf{Spin-wave maps for a $10^\circ$ interface.} Measurements are performed at an in-plane magnetic field of $B_{\mathrm{ip}} = 2.3~\mathrm{mT}$ and an NV excitation frequency of $f_{\mathrm{ESR}} = 2.8123~\mathrm{GHz}$. (\textbf{A}) Spatial map of the NV photoluminescence in the absence of microwave excitation. The light-grey region marks the YIG/NbTiN structure. (\textbf{B--D}) Spatial maps of the NV electron spin resonance contrast, $C_1$, acquired at different temperatures. } \label{fig:S16}
\end{figure*}

\begin{figure*}[!ht]
\includegraphics[width=\textwidth]{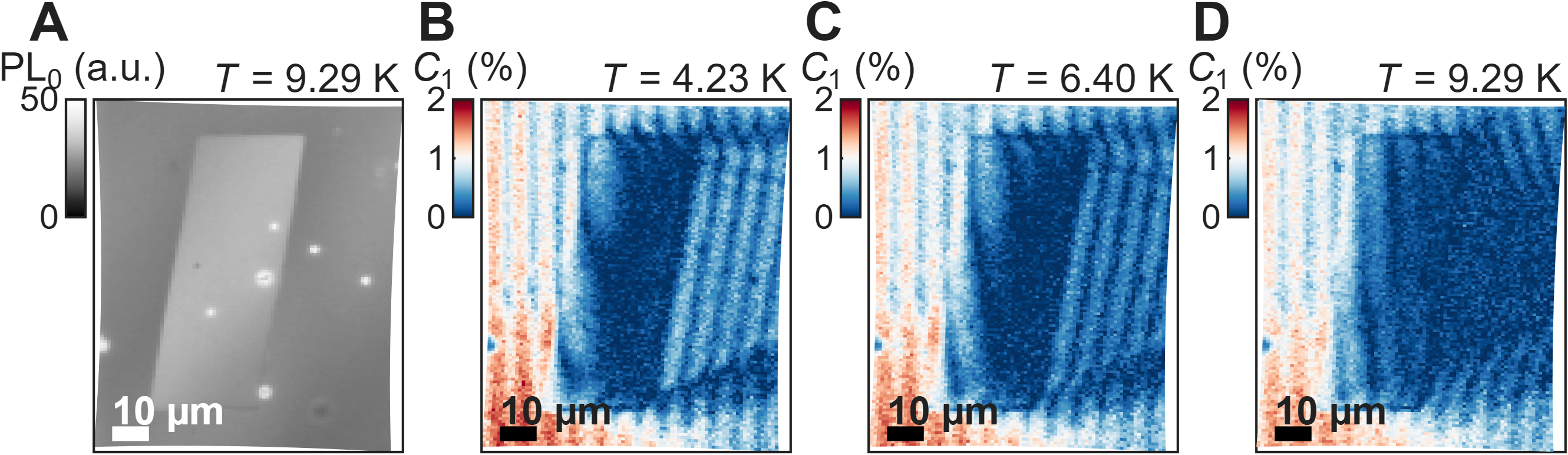}
\caption{ \textbf{Spin-wave maps for a $10^\circ$ interface.} Measurements are performed at an in-plane magnetic field of $B_{\mathrm{ip}} = 10.1~\mathrm{mT}$ and an NV excitation frequency of $f_{\mathrm{ESR}} = 2.5155~\mathrm{GHz}$. (\textbf{A}) Spatial map of the NV photoluminescence in the absence of microwave excitation. The light-grey region marks the YIG/NbTiN structure. (\textbf{B--D}) Spatial maps of the NV electron spin resonance contrast, $C_1$, acquired at different temperatures. } \label{fig:S17}
\end{figure*}

\begin{figure*}[!ht]
\includegraphics[width=\textwidth]{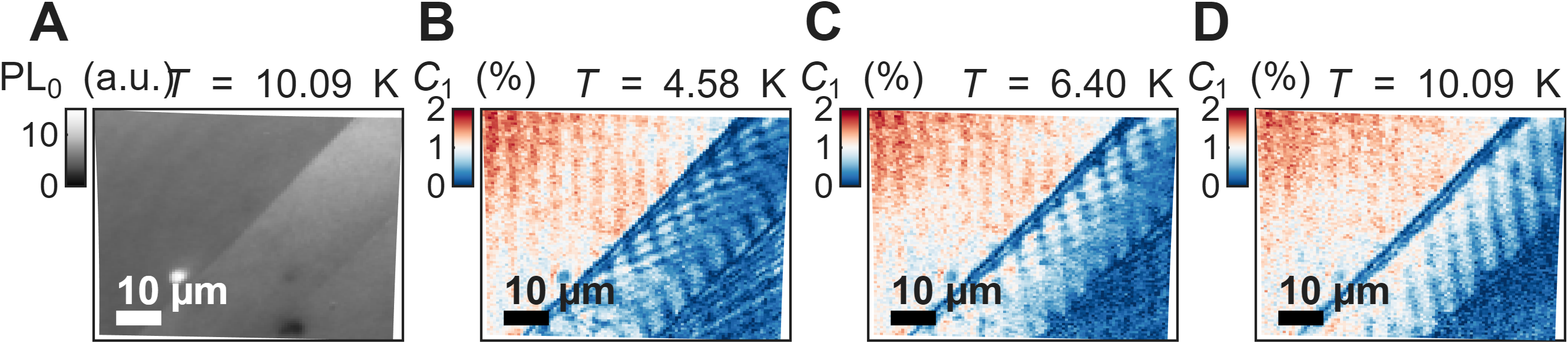}
\caption{ \textbf{Spin-wave maps for a $45^\circ$ interface.} Measurements are performed at an in-plane magnetic field of $B_{\mathrm{ip}} = 6.5~\mathrm{mT}$ and an NV excitation frequency of $f_{\mathrm{ESR}} = 2.6635~\mathrm{GHz}$. (\textbf{A}) Spatial map of the NV photoluminescence in the absence of microwave excitation. The light-grey region marks the YIG/NbTiN structure. (\textbf{B--D}) Spatial maps of the NV electron spin resonance contrast, $C_1$, acquired at different temperatures. } \label{fig:S18}
\end{figure*}

\begin{figure*}[!ht]
\includegraphics[width=\textwidth]{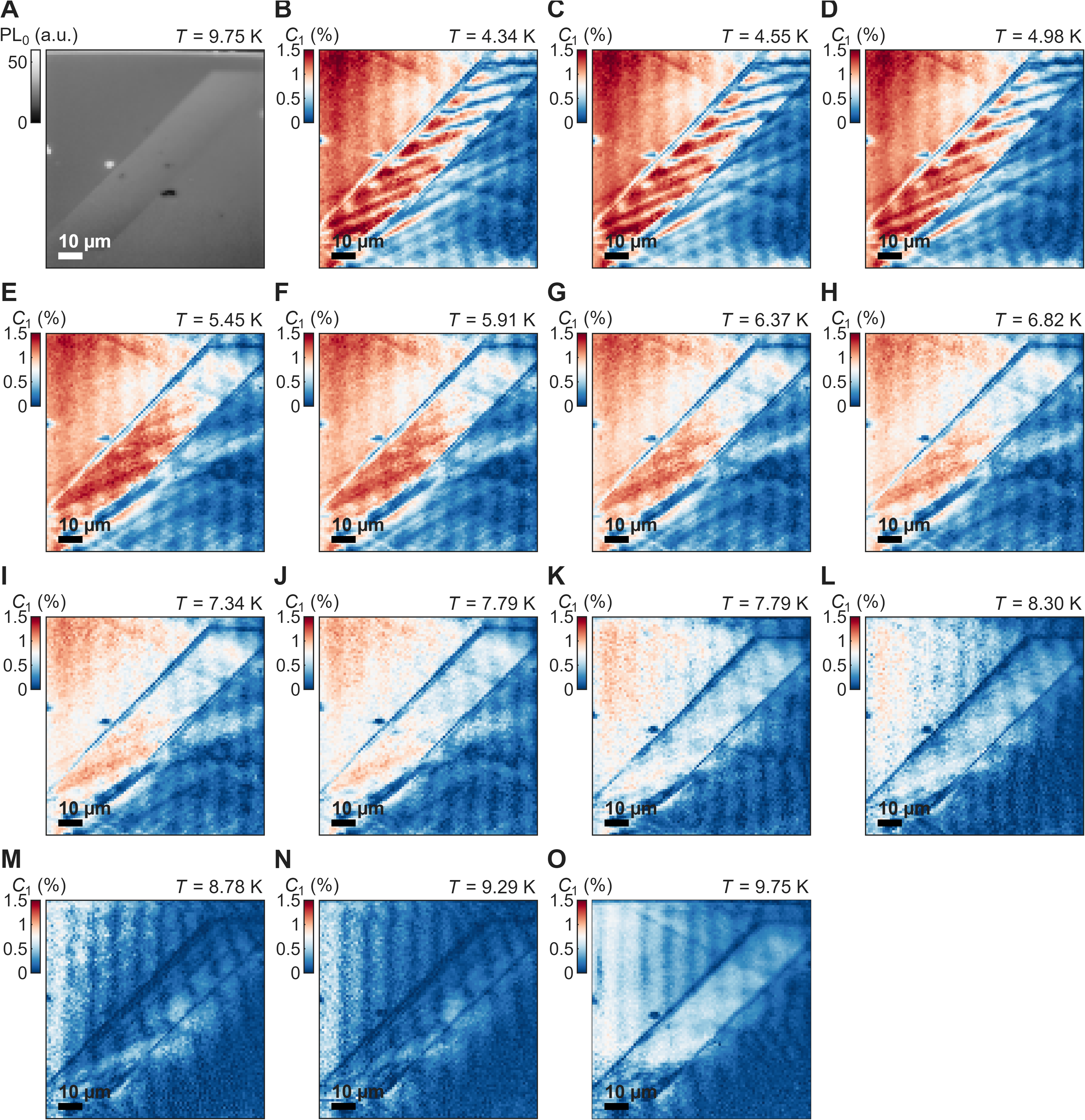}
\caption{ \textbf{Spin-wave maps for a $45^\circ$ interface.} Measurements are performed at an in-plane magnetic field of $B_{\mathrm{ip}} = 12.9~\mathrm{mT}$ and an NV excitation frequency of $f_{\mathrm{ESR}} = 2.5184~\mathrm{GHz}$. (\textbf{A}) Spatial map of the NV photoluminescence in the absence of microwave excitation. The light-grey region marks the YIG/NbTiN structure. (\textbf{B--D}) Spatial maps of the NV electron spin resonance contrast, $C_1$, acquired at different temperatures. } \label{fig:S19}
\end{figure*}

Additional spin-wave refraction maps were acquired for a range of magnetic fields and interface angles (Fig.~\ref{fig:S16}--\ref{fig:S19}). At low temperatures, these maps exhibit features consistent with those discussed above, including re-excitation at the second interface, signatures of negative group velocity, and the presence of additional spin-wave modes beneath the NbTiN layer.

\begin{figure*}[!ht]
\includegraphics[width=\textwidth]{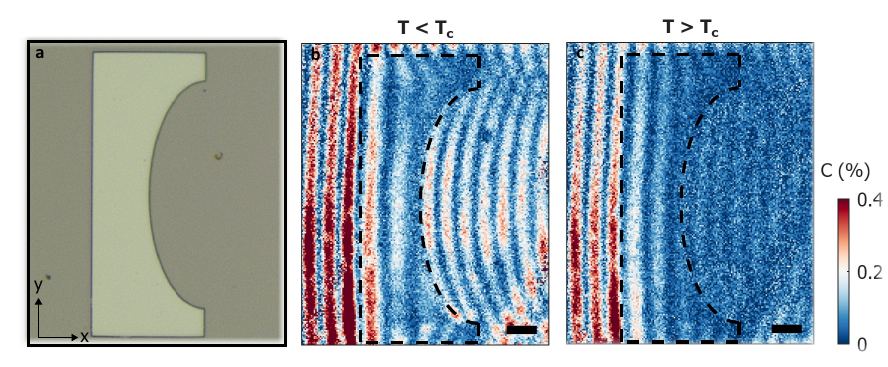}
\caption{ \textbf{Reconfigurable spin-wave phase front shaping using a lens-like superconducting structure.} (\textbf{A}) Optical microscopy image of the YIG film with a patterned NbTiN structure fabricated on top, acquired prior to placement of the diamond membrane. (\textbf{B, C}) Spatial maps of the NV electron spin resonance contrast recorded below (\textbf{B}) and above (\textbf{C}) the superconducting transition temperature. The scale bar corresponds to $10~\upmu\mathrm{m}$. Measurement parameters are $B_{\mathrm{ip}} = 10~\mathrm{mT}$ and $f_{\mathrm{ESR}} = 2.57~\mathrm{GHz}$. The temperatures are $T_{\mathrm{set}} = 5.9~\mathrm{K}$ for panel (\textbf{B}) and $T_{\mathrm{set}} = 11.3~\mathrm{K}$ for panel (\textbf{C}). } \label{fig:S20}
\end{figure*}

\begin{figure*}[!ht]
\includegraphics[width=\textwidth]{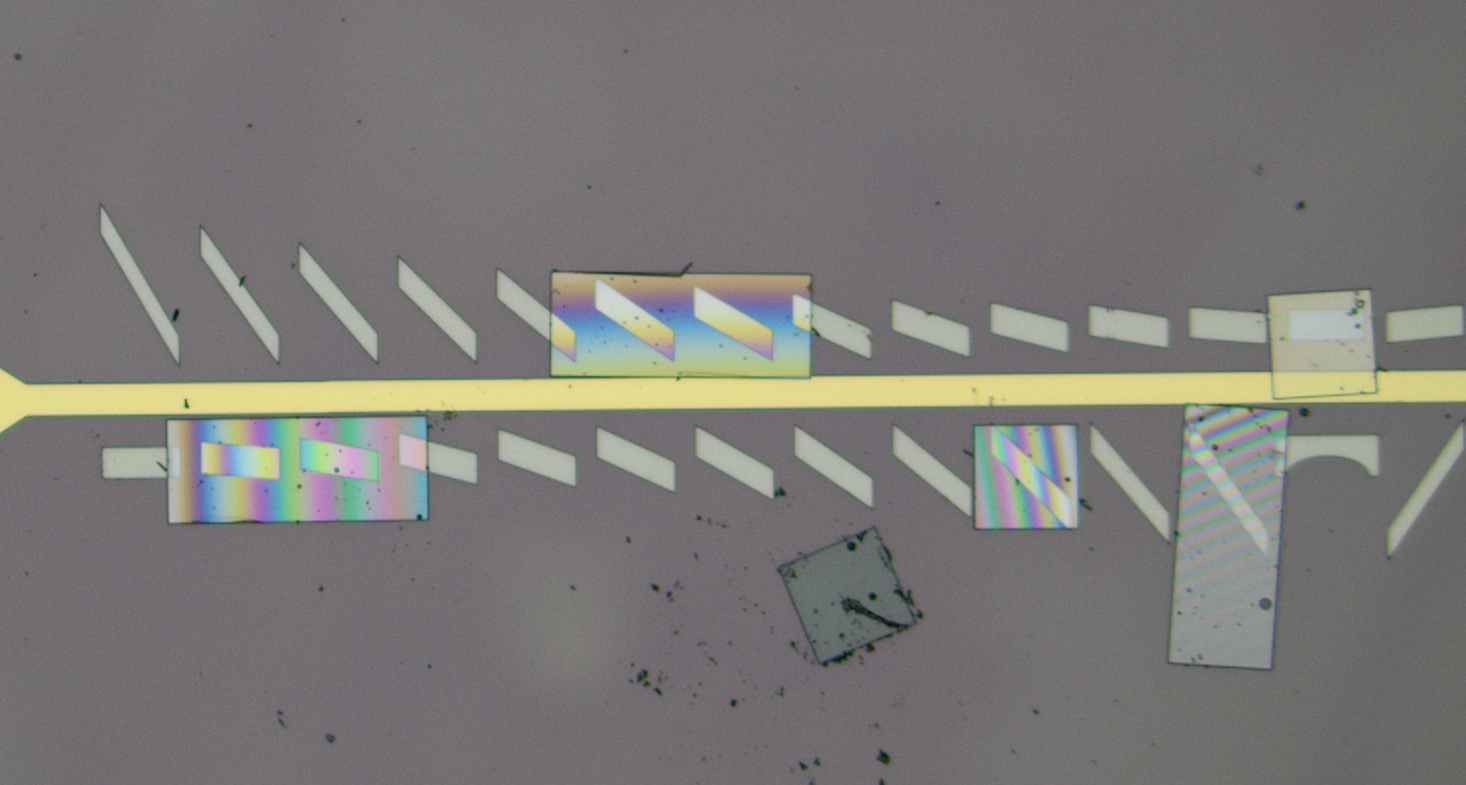}
\caption{ \textbf{NV-diamond membranes placed on superconducting elements.} Optical microscope image of $100 \times 100~\upmu\mathrm{m}^2$ and $250 \times 100~\upmu\mathrm{m}^2$ diamond membranes placed on NbTiN superconducting elements. } \label{fig:S21}
\end{figure*}

\clearpage

\begin{table}[H]
\centering
\begin{tabular}{llll}
\hline
parameter & & value & source \\
\hline
$M_s$ & saturation magnetization & $1.965(1)\cdot10^{5}\ \mathrm{A/m}$ & (VSM \cite{hansen_saturation_1974}) \\
$\gamma$ & gyromagnetic ratio & $2\pi\cdot28\ \mathrm{GHz/T}$ & (FMR-VNA \cite{serha_magnetic_2024}) \\
$t$ & YIG thickness & $204(1)\ \mathrm{nm}$ & manufacturer \\
$h$ & SC thickness & $150(1)\ \mathrm{nm}$ & sputtering \\
$D$ & exchange stiffness & $\sim1.51\cdot10^{-6}\ \mathrm{m^2/s}$ & \cite{noauthor_spin_2009} \\
$B_c$ & cubic anisotropy field & $12.2\pm0.8\ \mathrm{mT}$ & (FMR-VNA \cite{serha_magnetic_2024}) \\
$B_u^{\mathrm{YIG}}$ & uniaxial anisotropy field & $\sim\ \mathrm{mT}$ & fit \\
$B_u^{\mathrm{YIG/SC}}$ & uniaxial anisotropy field & $\sim\ \mathrm{mT}$ & fit \\
$\lambda_{\mathrm{L}}$ & London penetration depth & $\sim500\ \mathrm{nm}$ & fit \\
\hline
\end{tabular}

\caption{Material parameters used in the spin-wave dispersion model in the main manuscript.}
\label{tab:S1}
\end{table}

\clearpage
\bibliography{Refractionpaperv2}

%% file: sw_sc_abstract.tex
Spin waves are promising signal carriers for microwave control at the micrometer scale. However, realizing low-damping, tunable control of spin-wave propagation remains a central challenge. Here we use magnetic shielding by superconducting control elements to tune the local spin-wave dispersion and realize temperature-controlled refraction of spin waves in a thin-film magnetic insulator. Using magnetic imaging based on spins in diamond, we characterize the refractive index and demonstrate both positive and negative refraction as well as wavefront shaping by the superconductors. The observed refraction is explained by a geometrical analysis of the superconductivity-induced modification of the hyperbolic spin-wave dispersion. Our results demonstrate that superconductors enable tunable spin-wave optical elements, opening new opportunities for microwave control in classical or quantum information devices.